\documentclass[]{aa}  

\usepackage{graphicx}
\usepackage{txfonts}
\usepackage{amssymb}
\usepackage{multirow}
\usepackage{float}
\usepackage{gensymb}
\usepackage{xcolor}
\usepackage{verbatim} 
\usepackage{rotating}
\usepackage[]{hyperref}
\hypersetup{backref=true, pagebackref=true, hyperindex=true, breaklinks=true,colorlinks=true,urlcolor=blue, linkcolor=blue, citecolor=blue,pagecolor=red, bookmarks=true, bookmarksopen=true}

\begin{document}

   \title{ALMA chemical survey of disk-outflow sources in Taurus (ALMA-DOT)}

   \subtitle{V: Sample, overview, and demography of disk molecular emission}

   \author{A.\,Garufi \inst{\ref{Firenze}}
   \and L.\,Podio \inst{\ref{Firenze}}
   \and C.\,Codella \inst{\ref{Firenze}, \ref{IPAG}} 
   \and D.\,Fedele \inst{\ref{Firenze}}
   \and E.\,Bianchi \inst{\ref{IPAG}}
   \and C.\,Favre \inst{\ref{IPAG}}
   \and F.\,Bacciotti \inst{\ref{Firenze}}
   \and \\ C.\,Ceccarelli \inst{\ref{IPAG}}
   \and S.\,Mercimek \inst{\ref{Firenze}, \ref{UniFirenze}}
   \and K.\,Rygl \inst{\ref{IRA}}
   \and R.\,Teague \inst{\ref{CfA}}
   \and L.\,Testi\inst{\ref{ESO_Germany}, \ref{Firenze}}
    }

\institute{INAF, Osservatorio Astrofisico di Arcetri, Largo Enrico Fermi 5, I-50125 Firenze, Italy. \label{Firenze}
  \email{antonio.garufi@inaf.it}  
    \and Univ.\ Grenoble Alpes, CNRS, Institut de Plan\'{e}tologie et d'Astrophysique de Grenoble (IPAG), 38000 Grenoble, France \label{IPAG}
    \and{Universit\`{a} degli Studi di Firenze, Dipartimento di Fisica e Astronomia, Via G. Sansone 1, 50019 Sesto Fiorentino, Italy} \label{UniFirenze}
     \and INAF$-$Istituto di Radioastronomia \& Italian ALMA Regional Centre, via P.\,Gobetti 101, 40129 Bologna, Italy \label{IRA}
        \and Center for Astrophysics | Harvard \& Smithsonian, 60 Garden Street, Cambridge, MA 02138, United States of America \label{CfA}
    \and European Southern Observatory, Karl-Schwarzschild-Strasse 2, D-85748 Garching, Germany \label{ESO_Germany}
             }

   \date{Received -; accepted -}

 
 \abstract{We present an overview of the ALMA chemical survey of disk-outflow sources in Taurus (ALMA-DOT), a campaign devoted to the characterization of the molecular emission from partly embedded young stars. {The project is aimed at attaining at better understanding of the gaseous products delivered to planets by means of high-resolution maps of the assorted lines probing disks at the time of planet formation ($\lesssim1$ Myr).} {Nine} different molecules are surveyed through our observations of six Class I/flat-spectrum sources. As part of a series of articles analyzing specific targets and molecules, in this work we describe the sample and provide a general overview of the results, focusing specifically on the spatial distribution, {column densities, and abundance ratios of H$_2$CO, CS, and CN.} In these embedded sources, the $^{12}$CO emission is dominated by envelope and outflow emission while the CS {and, especially, the H$_2$CO} are good tracers of the gaseous disk structure. The spatial distribution and brightness of the o-H$_2$CO $3_{1,2}-2_{1,1}$ and CS 5$-$4 lines are very similar to each other and across all targets. The CN 2$-$1 line emission is fainter and distributed over radii larger than the dust continuum. The H$_2$CO and CS emission is always dimmed in the inner $\sim$50 au. While the {suppression} by the dusty disk and absorption by the line-of-sight material significantly contributes to this inner depression, an actual decrease in the column density is plausible in most cases, making the observed ring-like morphology realistic. We also found that the gaseous disk extent, when traced by H$_2$CO (150$-$390 au), is always 60\% larger than the dust disk. This systematic discrepancy may, in principle, be explained by the different optical depth of continuum and line emission without invoking any dust radial drift. Finally, the o-H$_2$CS $7_{1,6}-6_{1,5}$ and CH$_3$OH $5_{0,5}-4_{0,4}$ line emission are detected in two disks and one disk, respectively, while the HDO is never detected. {The H$_2$CO column densities are 12$-$50 times larger than those inferred for Class II sources while they are in line with those of other Class 0/I. The CS column densities are lower than those of H$_2$CO, which is an opposite trend with regard to Class II objects. We also} inferred abundance ratios between the various molecular species finding, among others, a H$_2$CS/H$_2$CO ratio that is systematically lower than unity (0.4$-$0.7 in HL Tau, 0.1$-0.2$ in IRAS 04302+2247, and <0.4 in all other sources), as well as a CH$_3$OH/H$_2$CO ratio (<0.7 in HL Tau and 0.5$-$0.7 in IRAS 04302+2247) that is lower than the only available estimate in {a protoplanetary disks} (1.3 in TW Hya) {and between one and two orders of magnitude lower than those of the hot corinos around Class 0 protostars. These results are a first step toward the characterization of the disk's chemical evolution, which ought to be complemented by subsequent observations of less exceptional disks and customized thermo-chemical modeling.}}


\authorrunning{Garufi et al.}

\titlerunning{ALMA-DOT V. Sample, overview, demography}

   \maketitle
%

\section{Introduction}
The chemical composition of protoplanetary disks affects the atmospheric composition of nascent planets. High angular- and spectral-resolution observations of protoplanetary disks are a stepping stone toward attaining an understanding of the formation pathway of increasingly complex molecules and their delivery to  planets undergoing formation. The Atacama Large Millimeter/submillimiter Array (ALMA) has provided the first highly resolved maps of the gas in protoplanetary disks \citep[e.g.,][]{Qi2013, Schwarz2016}. While most of the large surveys have, thus far, focused on carbon monoxide (CO) and its isotopologs \citep[e.g.,][]{Ansdell2016, Long2017}, given that { CO is the second-most abundant molecule after H$_2$}, some smaller surveys probing other molecules have been carried out \citep[e.g.,][]{Guzman2017, Bergner2018, Bergner2019, vantHoff2020}. At the same time, ALMA has enabled the detection of the first complex organic molecules (COMs), such as CH$_3$CN, CH$_3$OH, t-HCOOH, {and CH$_3$CHO \citep{Oberg2015, Walsh2016, Favre2018, vanthoff2018, Lee2019b}}. 

At present, the vast majority of protoplanetary disks with resolved molecular maps from ALMA are Class II sources, following the classification by \citet{Lada1987}, namely:\ objects older than $\sim$1 Myr, where the natal envelope has dissipated and the jets and outflows are less powerful with respect to the Class I phase. Younger sources are nonetheless fundamental toward the understanding of the planet formation. In fact, maps of the continuum emission from embedded sources have revealed the existence of disk substructures that may be related to the interaction with planets \citep[e.g.,][]{Sheehan2018, Fedele2018}. A prototypical example along this line is HL Tau \citep{ALMA2015}, a deeply embedded T Tau star (TTS) hosting a disk with multiple rings and gaps. A major limitation with regard to the molecular line observations of embedded Class I disks is prominent contamination on the part of the surrounding material, particularly when the given maps are tracing $^{12}$CO. This drawback supports the need for observations of other molecular species that are less affected by ambient contamination.

The ALMA chemical survey of disk-outflow sources in Taurus (ALMA-DOT) is a small campaign devoted to the characterization of the gas in young, embedded disks. The exquisite angular resolution offered by ALMA allowed us to separate the different spatial structures (disk, filaments, outflow, envelope) contributing to the strong molecular emission detected {on various scales around} these young Taurus sources \citep{Guilloteau2013}. The variety of the probed species is another fundamental aspect of the survey. As many as twenty-five spectral lines of {nine} molecular species have been probed. Beside CO, formaldehyde (H$_2$CO), carbon monosulfide (CS), and cyanide (CN) are the most characterized molecules in disks. These molecules have been imaged in a fair number of protoplanetary disks \citep[e.g.,][]{Loomis2015, Dutrey2017, Semenov2018, vanTerwisga2019}. The emission from these three molecules is often observed with a ring-like morphology \citep[and especially in old sources such as TW Hya,][]{Teague2016, Oberg2017}. However, recent surveys have revealed a variety of morphologies when the sample is diversified. \citet{Legal2019} and \citet{Pegues2020} showed both centrally peaked and \mbox{depressed} profiles for CS and H$_2$CO, respectively, with the occasional presence of outer rings and gaps. The other lines surveyed by ALMA-DOT are, to date, less characterized. Sulfur {monoxide (SO), sulfur} dioxide (SO$_2$ {and $^{34}$SO$_2$}), thioformaldehyde (H$_2$CS), and methanol (CH$_3$OH) have thus far been imaged in a handful of protoplanetary disks \citep{Semenov2018, Legal2019, Walsh2016} and deuterated water (HDO) has never been detected in disks.   

This paper is the fifth of the ALMA-DOT series and is devoted to the analysis of the {H$_2$CO, CS, and CN from the entire sample}. Accompanying papers focus on specific targets, {such as DG Tau \citep[Paper 0 and III,][]{Podio2019, Podio2020b}, DG Tau B \citep[Paper I,][]{Garufi2020b}, and IRAS 04302+2247 \citep[Paper II,][]{Podio2020}, or molecules, such as H$_2$CS \citep[Paper IV,][]{Codella2020} and SO$_2$ and SO (Paper VI, Garufi et al.\,in prep.).} This manuscript is organized as follows. In Sect.\,\ref{Survey_goal}, we illustrate the sample and goal of the survey. In Sect.\,\ref{Observations}, we describe the observing setup and the data reduction. In Sect.\,\ref{Results} we present the results of the analysis, and in Sects.\,\ref{Discussion} and \ref{Conclusions}, we discuss our findings and present our conclusions.    

\begin{table}
 \centering
 \caption{Stellar properties of all targets in the ALMA-DOT program.}
  \label{Star_properties}
  \begin{tabular}{lcccc}
  \hline
  Target & SpT & Mass (M$_{\odot}$) & $d$ (pc) & Ref. \\
  \hline
  DG Tau & K7 & 0.3 & 121.2 & a, 1 \\
  DG Tau B & - & 1.1 & 140 & b, 2 \\
  HL Tau & K3 & 2.1 & 147.3 & c, 3 \\
  Haro 6-13 & K5 & 1.0 & 130.4 & d, 1 \\
  IRAS 04302+2247 & - & 2.0 & 161 & e, 4 \\
  T Tau N, Sa, Sb & K0, -, M & 2, 2, 0.5 & 144.3 & f, 1 \\
   \hline
   \end{tabular}
   \tablefoot{Columns are: target name, spectral type, mass, and distance. Spectral types are, when available, from \citet{Herczeg2014}. References for the mass: a, \citet{Isella2009}; b, \citet{deValon2020}; c, \citet{Yen2019}; d, \citet{Schaefer2009}; e, \citet{Guilloteau2014}; f, \citet{Koehler2016}. References for the distance: 1, \citet{Gaia2018}; 2, \citet{Garufi2020b}; 3, \citet{Galli2018}; 4, based on the distance to the parent molecular cloud L1536 \citep{Galli2019}.}
\end{table}

\section{Survey goals and sample} \label{Survey_goal}
{The main goal of this survey is a comprehensive description of the molecular content of young ($\lesssim1$ Myr) disks and of the steps in the pathway of synthesis of increasingly complex molecules. The characterization of the chemical evolution from Class I to Class II disks of simple molecules beyond CO and, in particular, of simple organic molecules, such as H$_2$CO and CH$_3$OH, is an important step toward attaining an understanding of the initial conditions for planet formation and of the gaseous products that are eventually delivered to the planet body and atmospheres. For the scope of characterizing the gaseous structure of Class I disks, observation parameters must} spatially separate the disk emission from the emission of the extended envelope and the outflow. This goal requires us to resolve structures of a few hundred au, which translates, at the distance of Taurus, into an angular resolution on the order of 0.3\arcsec. The ALMA Band 6 allows us to probe one important transition of H$_2$CO and two of CH$_3$OH (see Sect.\,\ref{Observations} and Table \ref{Line_table}). The choice of the Band 6 also enables spectral windows (SPWs) centered on several hyperfine components of CN $N=2-1$ \citep[see][]{Guilloteau2013}, as well as of $^{12}$CO, $^{34}$SO$_2$, SO$_2$, and HDO. The broad SPW for continuum emission covers transitions of two additional species (CS and H$_2$CS), as well as another important transition of SO$_2$. Also, ALMA-DOT includes Band 5 observations {of DG Tau, which} enables the probing of five SO$_2$ lines and one SO line. Thus, as many as twenty-five transitions of {nine} different species were probed by the observations of the ALMA-DOT campaign, yielding one of the most spectrally varied imaging datasets that have ever been obtained for {circumstellar} disks. 

The sample of the ALMA-DOT campaign that we present in this work consists of six Taurus sources: DG Tau, DG Tau B, HL Tau, Haro 6-13, \mbox{IRAS 04302+2247} (hereafter IRAS04302), {and the triple system T Tau}. All these sources are partly embedded in the natal envelope and are known to drive prominent molecular outflows and atomic jets. The targets were selected from the single-dish-telescope survey by \citet{Guilloteau2013}, adopting the membership to Taurus as well as a bright continuum and integrated CN and H$_2$CO line fluxes as the primary criteria.  All these stars are TTSs, although their stellar masses are very diverse. DG Tau and the Sb member of the T Tau system are the only low-mass stars, while two targets are solar-mass stars (DG Tau B and Haro 6-13) and {four} are intermediate-mass stars (HL Tau, IRAS04302, T Tau N and Sa). The stellar properties are summarized in Table \ref{Star_properties}. 

All of these stars host very massive circumstellar disks (M$\rm _{dust}\sim50-270\ \rm M_{\oplus}$). From the Taurus sample by \citet{Guilloteau2013}, all our targets lie in the uppermost half of the dust-mass distribution with a three times larger average than the rest of the sample. Apart from that detail, our sample appears relatively varied. The extent of the dusty disk varies from $\sim$75 au (DG Tau) to $\sim$240 au (IRAS04302) and the disk inclination from nearly face-on (T Tau N) to nearly edge-on (IRAS04302). HL Tau is by far the most studied source being its disk a test-case for the ALMA high-resolution capabilities \citep{ALMA2015}. Besides the many studies on the disk structure \citep[e.g.,][]{Zhang2015, Pinte2016}, HL Tau has been recently studied in the context of the interaction between the disk and surrounding medium \citep[e.g.,][]{Yen2017, Wu2018}. DG Tau and DG Tau B are best studied for their prominent envelope and atomic jet \citep[e.g.,][]{Eisloffel1998, Bacciotti2000, Bacciotti2002, Coffey2007, Podio2011, deValon2020}. Their visual proximity may not be physical as their distance is possibly different \citep{Garufi2020b}. T Tau is a triple system \citep{Koresko2000} embedded in a complex environment of nebulosity and outflows \citep[e.g.,][]{vanBoekel2010, Kasper2016}. IRAS04302 has been named the butterfly star for its prominent bipolar cavity seen nearly face-on \citep{Lucas1997, Eisner2005}. Haro 6-13 is the least known of our targets, although a large disk has been mapped by \citet{Schaefer2009} with CO.

\begin{table}
 \centering
 \caption{Molecular lines probed by the ALMA-DOT campaign and analyzed in this work. The available dataset also includes several lines of SO and SO$_2$ that are described in a forthcoming work.}
 \label{Line_table}
  \begin{tabular}{ccccc}
  \hline
  Molecule & Transition & $\nu_{\rm rest}$ & E$\rm _{up}$ & S$_{ij}\mu^2$ \\
   & & (GHz) & (K) & (D$^2$) \\
  \hline
  \smallskip
  $^{12}$CO & 2$-$1 & 230.53800 & 17 & 0.02 \\
  \smallskip
  o-H$_2$CO & $3_{1,2}-2_{1,1}$ & 225.69777 & 33 & 43.5 \\
  \smallskip
  CS & 5$-$4 & 244.93555 & 35 & 19.1 \\
  \smallskip
  \multirow{11}{*}{CN} & 2$-$1, $\frac{3}{2}-\frac{1}{2}$, $\frac{3}{2}-\frac{3}{2}$ & 226.63219 & 16 & 1.2 \\
  \smallskip
  & 2$-$1, $\frac{3}{2}-\frac{1}{2}$, $\frac{5}{2}-\frac{3}{2}$ & 226.65956 & 16 & 4.2 \\
  \smallskip
  & 2$-$1, $\frac{3}{2}-\frac{1}{2}$, $\frac{1}{2}-\frac{1}{2}$ & 226.66369$^b$ & 16 & 1.2 \\
  \smallskip
  & 2$-$1, $\frac{3}{2}-\frac{1}{2}$, $\frac{3}{2}-\frac{1}{2}$ & 226.67931 & 16 & 1.6 \\
  \smallskip 
   & 2$-$1, $\frac{5}{2}-\frac{3}{2}$, $\frac{7}{2}-\frac{5}{2}$ & 226.87478 & 16 & 6.7 \\
  \smallskip 
   & 2$-$1, $\frac{5}{2}-\frac{3}{2}$, $\frac{5}{2}-\frac{3}{2}$ & 226.87419$^b$ & 16 & 4.2 \\
   \smallskip
   & 2$-$1, $\frac{5}{2}-\frac{3}{2}$, $\frac{3}{2}-\frac{1}{2}$ & 226.87589$^b$ & 16 & 2.6 \\
   \smallskip
   & 2$-$1, $\frac{5}{2}-\frac{3}{2}$, $\frac{3}{2}-\frac{3}{2}$ & 226.88742 & 16 & 0.8 \\
   \smallskip
   & 2$-$1, $\frac{5}{2}-\frac{3}{2}$, $\frac{5}{2}-\frac{5}{2}$ & 226.89212 & 16 & 0.8 \\
   \smallskip
    \multirow{2}{*}{HDO} & $3_{1,2}-2_{2,1}$ & 225.89672 & 168 & 0.7 \\
    \smallskip
     & $2_{1,1}-2_{1,2}$ & 241.56155 & 95 & 0.4 \\
   \multirow{2}{*}{CH$_3$OH} & $3_{-2,2}-4_{-1,4}$ (E) & 230.02706 & 40 & 2.9 \\
   \smallskip
     & $5_{0,5}-4_{0,4}$ (A) & 241.79143 & 35 & 16.2 \\
     o-H$_2$CS & $7_{1,6}-6_{1,5}$ & 244.04850 & 60 & 55.9 \\
   \hline
   \end{tabular}
   \tablefoot{Columns: molecular species, transition, frequency at rest frame, upper-level energy, and line strength. The label $^b$ indicates that the line is blended with the previous line at the present kinematical width and observing spectral resolution. All parameters are from CDMS \citep{Mueller2005}.}
\end{table}

\section{Observations and data reduction} \label{Observations}
In this work, we make use of ALMA observations performed in Cycle 4 and Cycle 6. The Cycle 4 observations of DG Tau and DG Tau B taken in August 2017 were presented by \citet{Podio2019} and \citet{Garufi2020b}, respectively. {The maps of this observing run have a beam size of the order of 0.11\arcsec$-$0.17\arcsec, a maximum recoverable scale of $\sim1.3\arcsec$, and an rms\ of 0.6$-$1.3 mJy beam$^{-1}$ (see reference papers and Table \ref{Integrated_fluxes_1} of this work).}

The Cycle 6 observations of HL Tau, T Tau, \mbox{IRAS04302}, and Haro 6-13 were obtained on October 28, 2018 in an extended configuration with baselines ranging from 15 m to 1.4 km. The largest angular scale is $\sim$5\arcsec\ ($\sim$700 au). The total integration time spans from $\sim$104 to $\sim$113 minutes. The bandpass calibration exploited the quasar J0423-0120 and the phase calibration the quasar J0510+1800. The correlator setup included 12 SPWs with 0.244 MHz resolution. The relevant molecular transitions sitting within these SPWs are shown in Table \ref{Line_table}. The molecular parameters are taken from the Cologne Database of Molecular Spectroscopy (CDMS) \citep{Mueller2005}.

The data reduction followed the standard procedure using the Common Astronomy Software Applications package \citep[CASA,][]{McMullin2007} version 4.7.2. Self-calibration was carried out through the continuum of each source by applying the solutions in phase to the line-free continuum and continuum-subtracted SPWs. Since each continuum emission is strong, solution intervals of the self-calibration could be decreased down to 10 sec without rejecting any solution. This operation improved the S/N of the continuum image of HL Tau, T Tau, \mbox{IRAS04302}, and Haro 6-13 by a factor of 3.3, 4.5, 3.4, and 1.2, respectively. The same self-calibration table was applied to all SPWs. Continuum images and spectral cubes were produced using \textsc{tclean} interactively, by applying a manually selected mask on the visible signal and iterating until the visual inspection of the residuals revealed no significant source emission. The maps presented in this work are obtained using Briggs weighting of 0.0 and setting a channel width of 0.2 km s$^{-1}$. Alternatively, the maps of the faint CH$_3$OH and SO$_2$ lines analyzed by \citet{Podio2020} and Garufi et al.\,(in prep.)\ are obtained with weighting of 2.0 to maximize the recoverable flux. The clean beam size {of the 2018 observing run spans from 0.26\arcsec\ to 0.34\arcsec\ and the rms\ of the resulting line cubes from 0.9 to 2.2 mJy beam$^{-1}$} (see Table \ref{Integrated_fluxes_1}).

\begin{figure*}
  \centering
 \includegraphics[width=18cm]{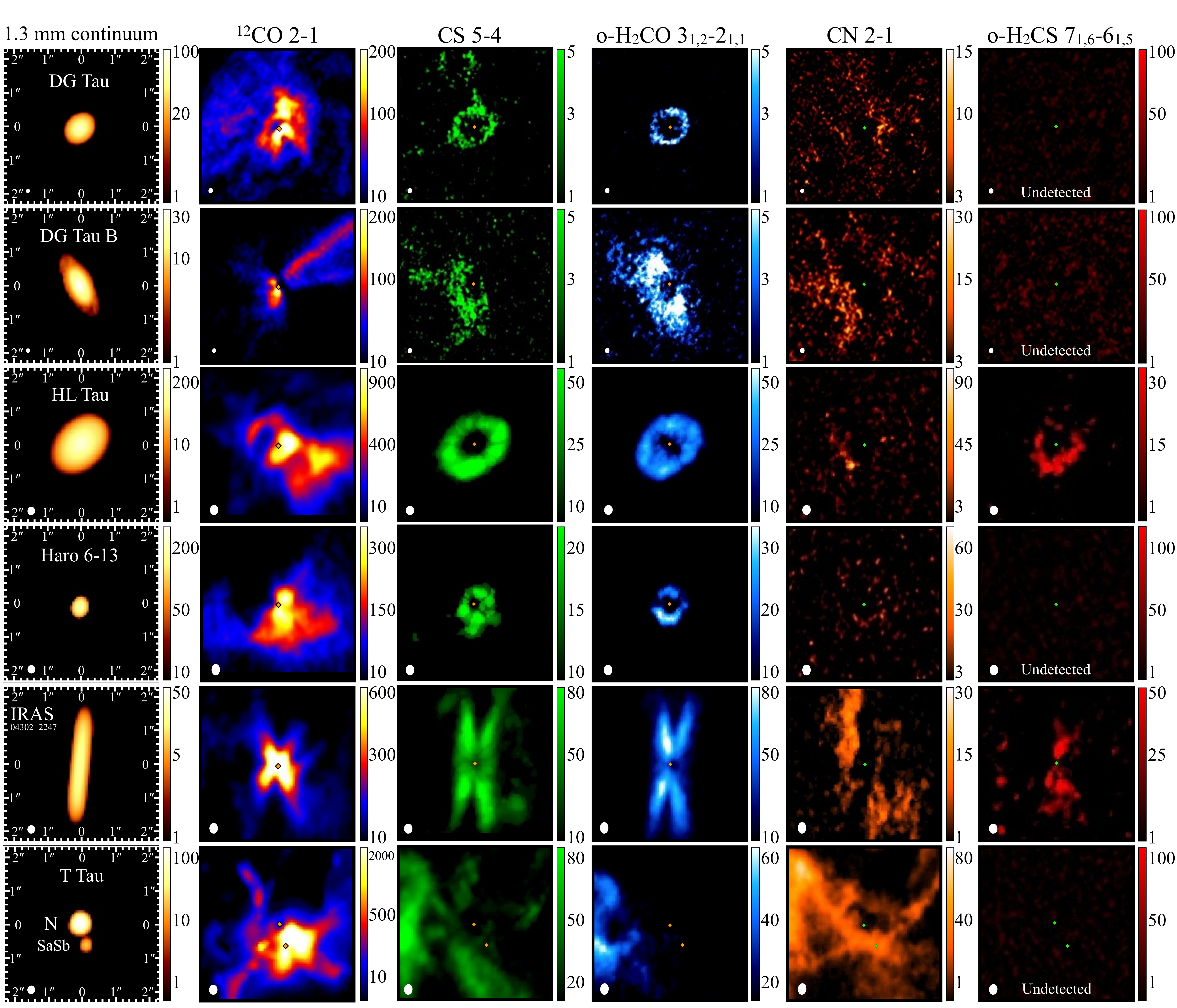} 
     \caption{{Morphological} overview of the sample. Continuum and molecular emission from $^{12}$CO, CS, H$_2$CO, CN, and H$_2$CS is shown for the whole sample. Images of spectral lines are the flux-integrated map (moment 0) obtained by clipping fluxes below 3$\sigma$ (the CO, CS, and H$_2$CO), 2$\sigma$ (the CN and detected H$_2$CS), or without any clip (the undetected H$_2$CS). Continuum fluxes are expressed in mJy beam$^{-1}$ while line fluxes in mJy beam$^{-1}$ km s$^{-1}$. The {symbols} in the center of the images indicate the geometrical center of the continuum emission. The beam size is indicated to the bottom-left corner of each panel. All images have the same angular scale, as indicated in the leftmost column. North is up, east is left.} 
 \label{Overview}
 \end{figure*}

\section{Results} \label{Results}

\subsection{Overview of ALMA-DOT results}
The main {emission characteristics inferred by} the ALMA-DOT campaign can be summarized as follows:

 The continuum, CO, CS, H$_2$CO, and CN emission are always detected {(although not all CN hyperfine lines are)}. The CO emission evidently originates from both the disk and the surrounding medium. With one notable exception (T Tau), the CS, H$_2$CO, and CN emission primarily come from the disk.

The spatial distribution of the H$_2$CO $3_{1,2}-2_{1,1}$ and CS 5$-$4 line emission is similar, whereas that of the CN 2$-$1 line significantly differs. The flux from each of these three molecules is dimmed in the center {(this work)}.

 The H$_2$CS emission is detected in two sources {(HL Tau and IRAS04302)}. This signal clearly originates from the disk and its morphology resembles that of CS and H$_2$CO \citep[see][]{Codella2020}.

 The SO$_2$ emission is detected in four sources (DG Tau, HL Tau, IRAS04302, and T Tau) {and SO is detected in the only source (DG Tau) where it is probed}. This signal is associated with the interaction between the disk and the surrounding medium since it is found in correspondence with extended filaments seen in CO and CN (see Garufi et al.\,in prep.).

The CH$_3$OH emission is {tentatively} detected in one source \citep[IRAS04302, see][]{Podio2020}. The HDO emission is never detected. Prominent outflow cones are visible around DG Tau B \citep[see][]{Garufi2020b}. {Similar structures around HL Tau are to be described by Bacciotti et al.\,(in prep.).}

\begin{figure*}
  \centering
 \includegraphics[width=18cm]{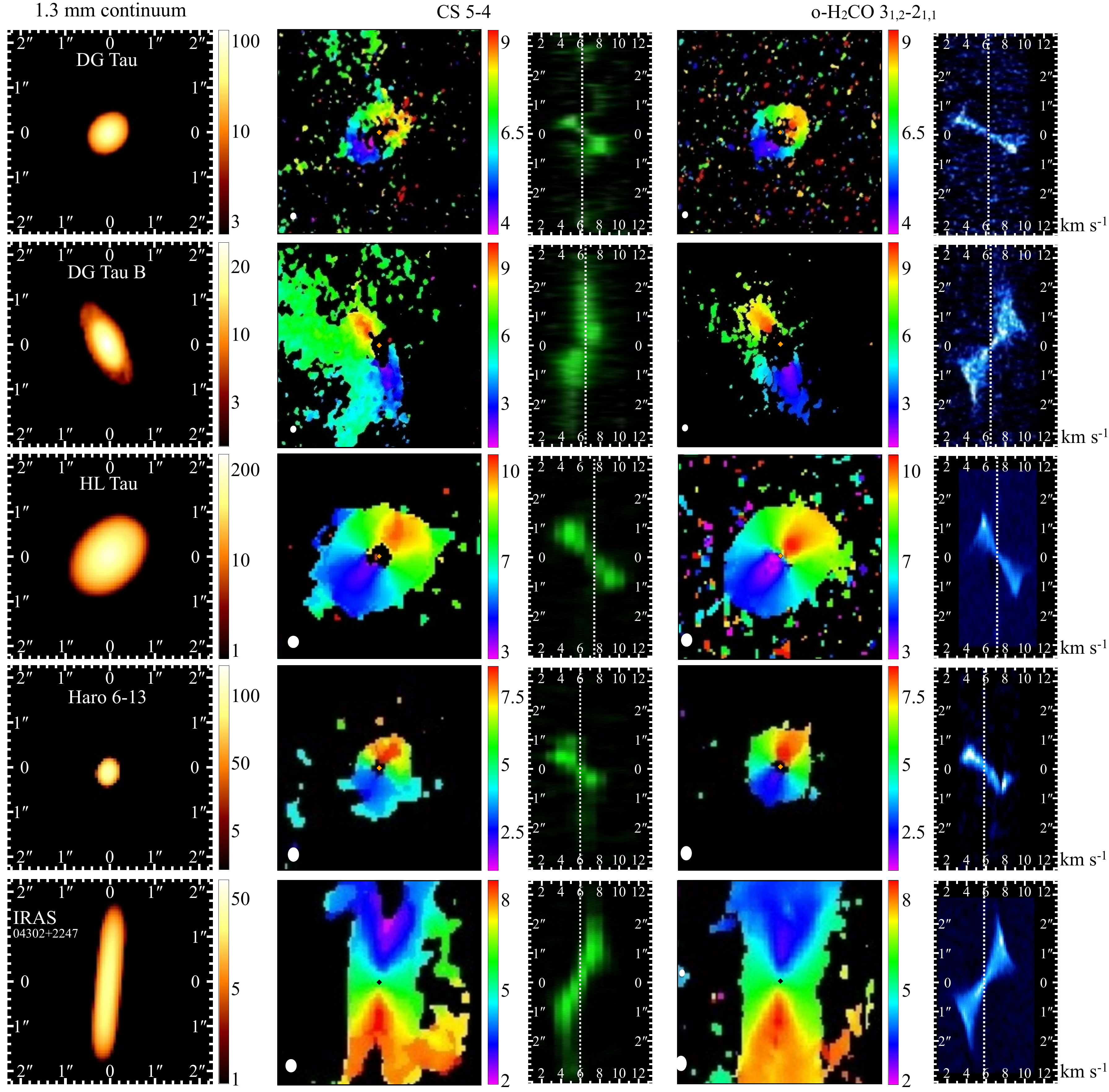}
     \caption{{Kinematical overview of the sample. The intensity-weighted velocity maps (moment 1, square panels) and position-velocity diagrams (PV, vertical panels) of the CS and H$_2$CO molecular emission are shown for the sample analyzed in this work along with the continuum emission of Fig.\,\ref{Overview}. Moment 1 maps are obtained by clipping fluxes below 3$\sigma$. PV diagrams are obtained over the disk major axis taking a width as large as the projected gaseous disk size (see Table \ref{Disk_properties}) along the minor axis. The color units of the moment 1 maps are in km s$^{-1}$, those of the PV diagrams are the same of Fig.\,\ref{Overview}. {The x-axis of the PV diagrams is in km s$^{-1}$.} The symbols in the center of the images indicate the geometrical center of the continuum emission. The beam size is indicated to the bottom-left corner of each panel. The vertical dashed lines indicate the systemic velocities of the system (see Appendix \ref{Spectral_profile}). All moment 0 and 1 maps have the same angular scale, as indicated in the leftmost column. North is up, east is left.}} 
 \label{Overview_kinematics}
 \end{figure*}


This work is devoted to {the molecular emission associated with the disk, namely, the CS, CN, H$_2$CO, and H$_2$CS emission from all sources except T Tau. An analysis of T Tau and of the SO and SO$_2$ lines is to be presented in a forthcoming paper.} Figure \ref{Overview} gives a {morphological} overview of the continuum and molecular line emission {analyzed in this work, while Fig.\,\ref{Overview_kinematics} gives a kinematical overview of the CS and H$_2$CO line emission}. The individual cases are described in Sect.\,\ref{Individual_sources}. The spatial characterization of the bright lines of CS, CN, and H$_2$CO is presented in Sect.\,\ref{Spatial_distribution} and the calculation of the column density of all disk molecular species is given in Sect.\,\ref{Line_brightness}. Out of the many CN hyperfine lines probed, only the 226.87478 GHz line is used to constrain the spatial distribution since this line is significantly brighter. 

\subsection{Individual sources} \label{Individual_sources}
An in-depth analysis of some individual targets has been provided in a series of accompanying papers \citep[Bacciotti et al.\,in prep.]{Podio2019, Podio2020, Podio2020b, Garufi2020b}. Here, we give a brief description of each source to pave the way for the demographical analysis of Sect.\,\ref{Spatial_distribution}.

\textit{DG Tau}: Both CS and H$_2$CO lines are emitted in a ring-like structure located at the outer edge of the dust continuum emission \citep{Podio2019, Podio2020b}. The CN emission is significantly fainter and appears to originate from larger separations. The inner region shows null values {(the H$_2$CO)} or marginally negative fluxes {(CS and CN)}, after the process of continuum subtraction (see Sect.\,\ref{Observations}).

\textit{DG Tau B}: Analogously to DG Tau, the emission from CS and H$_2$CO is co-spatial whereas that of CN is fainter and located at larger radii. {Both CS and H$_2$CO appear asymmetrically distributed along the minor axis because of the disk inclination} \citep[see][]{Garufi2020b}. Unlike DG Tau, the inner region shows severely negative fluxes. This region coincides with the separation over which the continuum emission at 1.3 mm is optically thick \citep{Garufi2020b}. 

\textit{HL Tau}: A ring in both CS and H$_2$CO is visible in the respective map. As for the other sources, the CN emission is fainter and more diffuse. Although the ring-like structure is reminiscent of DG Tau, the presence of severely negative values in the inner region is similar to the case of DG Tau B. H$_2$CS is also detected from a ring-like structure. Finally, two prominent outflow cones, similar to those of DG Tau B, are detected in CO.

\textit{Haro 6-13}: This target hosts the smallest and faintest disk of the sample. Nevertheless, two prominent outflow cones are visible in CO. The distribution of all gaseous lines from the disk resembles what is seen in DG Tau. 

\textit{IRAS04302}. Having the disk {at} an orientation close to 90$\degree$, this target allows to study the vertical extent of the molecular emission rather than the radial distribution \citep[see][]{Podio2020}. CO, CS, and H$_2$CO lines are all distributed in a X shape, while CN seems to be emitted from two asymmetric regions running parallel to the disk extension. 

\textit{T Tau}: This is the only source in our sample where the CS, H$_2$CO, and CN emission traces filamentary structures around the two stars and not the disk. The absence of any signal from T Tau N may be explained by the face-on geometry of its disk, which provides gas emission at the systemic velocity only, where the contaminating flux from the ambient gas is strong. The absence of any obvious signal from the T Tau Sa and Sb components is likely ascribed to the bright filaments lying close to these stars. This source is therefore the subject of another study set in the context of the interaction between disk and medium (Garufi et al.\,in prep.) and, thus, it is no longer discussed in this work.

\subsection{Spatial distribution of detectable emission} \label{Spatial_distribution}
While the CO emission from the disk is evidently contaminated by the surrounding emission (see Fig.\,\ref{Overview}), the CS, CN, and H$_2$CO emission from the inner $\sim$200 au appear to be related to the gaseous disk structure. In this section, we focus on the emission from these three molecules. {The CS and H$_2$CO line emission is bright enough to enable a meaningful kinematical investigation. Both the intensity-weighted velocity maps (moment 1) and position-velocity (PV) diagrams, shown in Fig.\,\ref{Overview_kinematics}, reveal that their emission is primarily (for the CS) and almost entirely (for the H$_2$CO) originated in the disk. Some envelope or outflow contamination is seen in CS as rest-frame emission at radii larger than the main emission with a Keplerian pattern (see in particular DG Tau, DG Tau B, and Haro 6-13). However, none of these features is seen in H$_2$CO from the moment 1 maps nor from the PV diagrams, suggesting that this molecule is an optimal tracer of the gaseous disk structure in these sources. The faint flux of CN and the presence of lines blended with the main line (see Table \ref{Line_table}) make the kinematical investigation of the CN emission impracticable.} 

\begin{table}
 \centering
 \caption{Properties of dust and H$_2$CO emission.}
  \label{Disk_properties}
  \begin{tabular}{lccccccc}
  \hline
  Target & P.A. & $i$ & $z/r$ & $\rm r_{out, d}$ & $\rm r_{in,g}$ & $\rm r_{out,g}$ & $\rm \delta\, r_{out}$ \\
   & ($\degree$) & ($\degree$) &  & (au) & (au) & \\
  \hline
  DG Tau & 135 & 35 & 0.09 & 75 & 35 & 120 & 1.6 \\
  DG Tau B & 24 & 62 & 0.21 & 150 & 45 & 240 & 1.6 \\
  HL Tau & 138 & 47 & 0.16 & 150 & 55 & 250 & 1.7 \\
  Haro 6-13 & 167 & 38 & 0.07 & 100 & 30 & 150 & 1.5  \\
  IRAS 04302 & 175 & 81 & 0.23 & 240 & 50 & 390  & 1.6 \\
   \hline
   \end{tabular}
   \tablefoot{Columns: target name, disk position angle, inclination, opening angle, outer radius of the dust emission, inner and outer radius of the gas emission, and ratio between the dust and gas outer radii.}
\end{table}

\begin{figure}
  \centering
 \includegraphics[width=9cm]{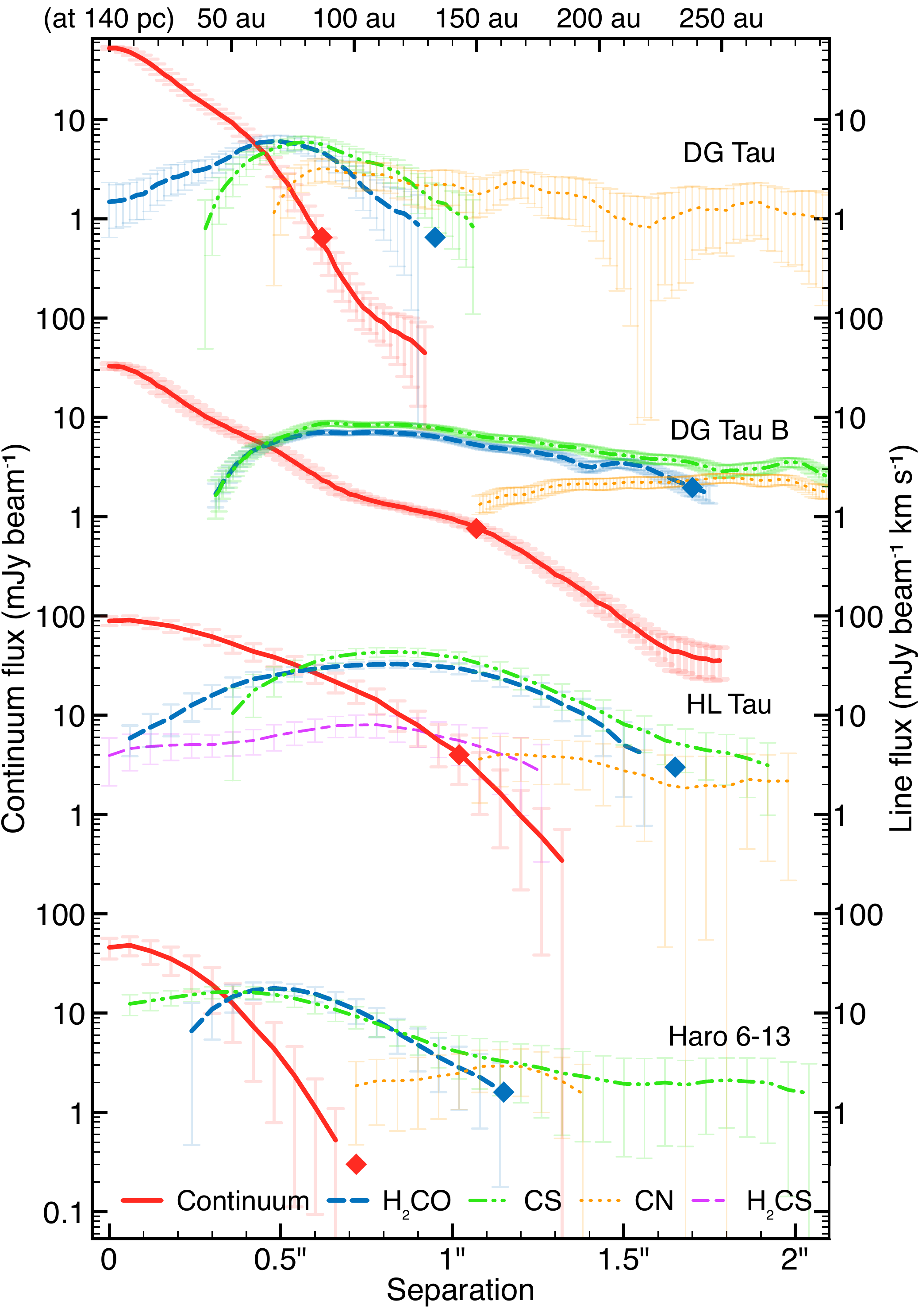} 
     \caption{Radial profile of the continuum and spectral line emission, obtained from the azimuthal average as described in Sect.\,\ref{Radial_distribution}. A vertical offset among objects has been put for better visualization. Error bars are obtained by propagating the uncertainties at all angles and do not include the dispersion of the averaged sample. Only fluxes above 3$\sigma$ confidence are shown. {The {red and blue} diamonds indicate the dust and gas outer radii constrained from the 90\%-flux method (see text)}. The physical scale shown at top are indicative for a distance of 140 pc, although sources are spread around this value (see Table \ref{Star_properties}).} 
 \label{Radial_profile}
 \end{figure}

\subsubsection{Vertical distribution} \label{Vertical_profile}
{Much of} the molecular emission {of circumstellar disks is thought to} originate from an intermediate layer between mid-plane and disk atmosphere {\citep[see e.g.,][]{Teague2020, Podio2020}}. The vertical distribution of the emission also alters the appearance of its radial distribution if the disk is even only moderately inclined. We employed two methods to constrain the vertical origin of the molecular emission. The first, which is by far more reliable, could only be adopted for the edge-on disk of IRAS04302, where we directly measured the opening angle between the midplane and the height $z$ with maximized H$_2$CO flux \citep[see][]{Podio2020}, obtaining $z/r=0.21 - 0.25$. For the other disks, we could only constrain $z/r$ from the apparent shift of apparent substructures along the minor axis. This method is routinely adopted for near-IR observations \citep[e.g.,][]{deBoer2016}. Assuming that all apparent substructures are azimuthally symmetric, we applied this method to the visible rings in H$_2$CO {and CS} of HL Tau, Haro 6-13 and DG Tau, as well as to the sharp outer edge of the main emission in DG Tau B \citep[see also][]{Garufi2020b}. First, we constrained the disk centroid, position angle (PA), and inclination, $i,$ with a Gaussian fit to the continuum emission. The results are consistent with those obtained from higher resolution images such as that of HL Tau \citep{ALMA2015}. {Then we generated the ellipse expected from the projection of a circular ring with the measured PA\ and $i$. The offset of this ellipse along the minor axis increases as we increase the $z/r$ and this is done until we visually fit the observed substructure with the synthetic ellipse.} 

The results of this exercise are shown in Table \ref{Disk_properties}. {We only list the values obtained from H$_2$CO since they are nearly identical to those from CS.} \mbox{Also, DG Tau B} and HL Tau show a larger value of $z/r$, as is intuitive from the observed asymmetry between near and far disk sides. {These values are comparable with that of IRAS04302, making the opening angle of the largest disks in our sample analogous}. We however caution that the moderate angular resolution, the small size of the rings (in particular in Haro 6-13), and the assumption of azimuthally symmetric structures \citep[which may not hold in DG Tau, see][]{Podio2019} limit the accuracy of our constraints. These estimates are nonetheless fundamental in view of the extraction of the radial profile. {In fact, this can shift up to 0.2\arcsec\ if the disk opening angle is not taken into account, while only a minor shift is yielded by the range of uncertainties on the $z/r$ that we constrained.}


\subsubsection{Radial distribution} \label{Radial_distribution}
As is clear from Fig.\,\ref{Overview}, four targets can be used to probe the radial distribution of the line emission from the disk. The radial profiles of DG Tau, DG Tau B, HL Tau, and Haro 6-13 are shown in Fig.\,\ref{Radial_profile}. All profiles are obtained by azimuthally averaging the emission over annuli with the PA, $i$, and $z/r$ that are constrained in Sect.\,\ref{Vertical_profile} and shown in Table \ref{Disk_properties}.

The radial profiles of Fig.\,\ref{Radial_profile} show some major analogies and a few differences. First of all, the gaseous emission is always more extended than the dust emission. To alleviate the dependence on the signal sensitivity (which is very different for continuum and molecular emission), we defined the outer radius of a certain type of emission as the separation from the star enclosing 90\% of the total flux in the image. To constrain the gas outer radius, we used the H$_2$CO emission since this line is the least affected by the surrounding envelope (see Sect.\,\ref{Outer_disk}). The resulting outer radii for the dust and gas emission are shown in Table \ref{Disk_properties}. Disk sizes in the sample are very diverse but the ratios between the measurements from these two components are surprisingly similar (1.6 with a scatter of less than 0.1). This finding is discussed in Sect.\,\ref{Outer_disk}.

The two sources observed at high angular resolution ($\theta \sim$ 0.15\arcsec), DG Tau and DG Tau B, show dust substructures \citep[see][]{Podio2019, Garufi2020b}, while the other two ($\theta \sim$ 0.3\arcsec) do not. Clearly, this resolution is insufficient to resolve dust substructures as fine as those seen in HL Tau \citep{ALMA2015} and that can be present in all disks although unresolved. The gas radial profiles also do not show any evidence of substructures apart from the main rings over which most of the flux is emitted. Between approximately 50 and 150 au, the CS and H$_2$CO curves are very similar in both brightness and trend. However, the trends diverge further out since the CS emission extends over slightly larger radii. The CN flux is nearly one order of magnitude fainter and the distribution does not show any obvious analogy with those of the other molecules.

 \begin{figure}
  \centering
 \includegraphics[width=8.8cm]{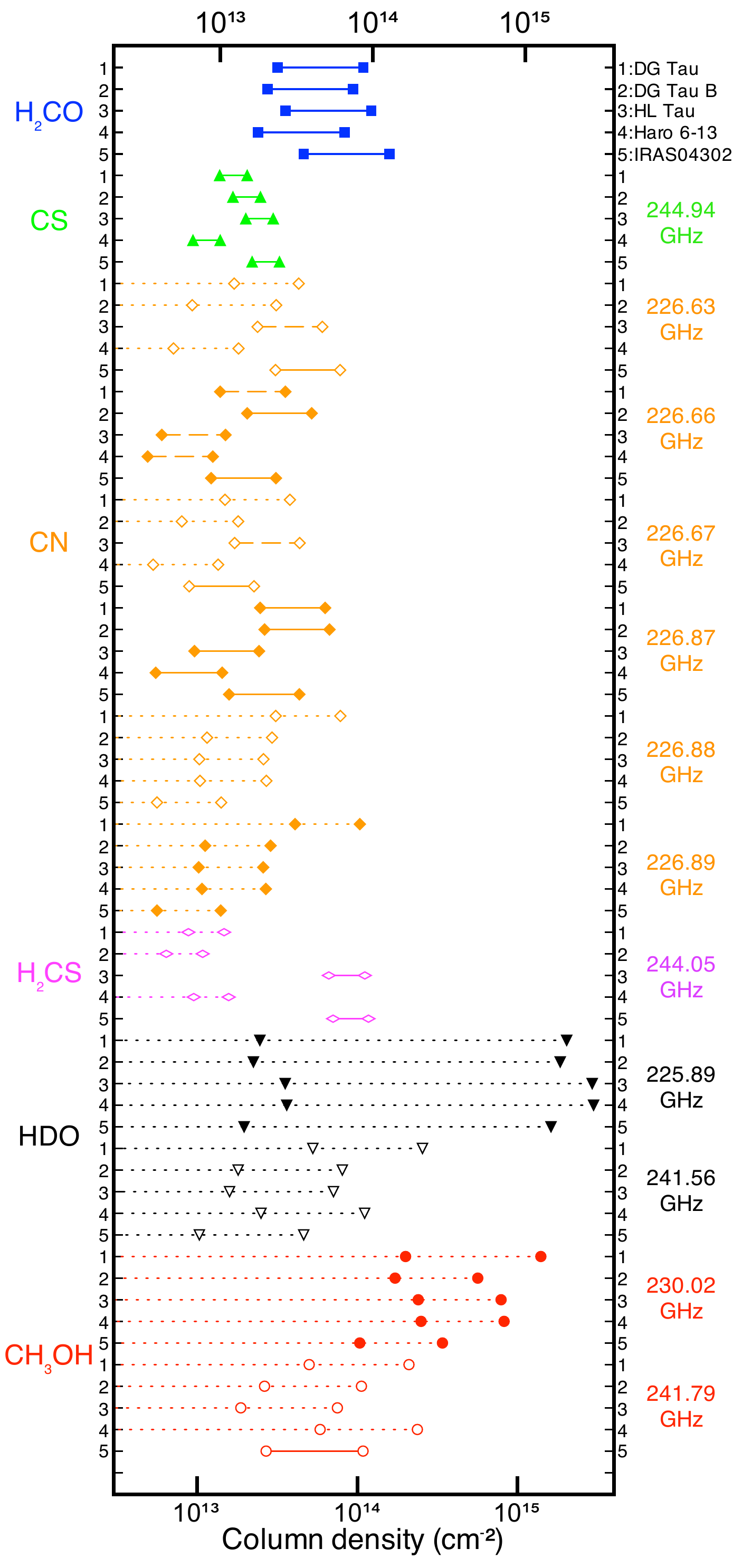} 
     \caption{Column densities of all molecules calculated in the hypothesis of optically thin lines and LTE conditions. The range of solutions is obtained over the relative emitting region only and reflects the minimum and maximum values found over a range of temperatures between 20 K and 100 K. Dotted lines indicate the upper limits derived from the undetected lines. {Dashed lines indicate column densities constrained from 2$\sigma$ detections. Filled and empty symbols are drawn for an easier reading of different lines.} } 
 \label{Column_densities}
 \end{figure}

Finally, all lines show a diminished flux toward the center. The inner onset of CN always occurs at larger radii while those of CS and H$_2$CO are, broadly speaking, comparable. More specifically, the two inner trends are identical in DG Tau B while H$_2$CO is found at slightly smaller separations in DG Tau and HL Tau but at slightly larger separations in Haro 6-13. Over this inner region, the fluxes are  frequently found to be negative around the systemic velocity $V_{\rm sys}$. This effect can be appreciated from the spectral profiles \citep[see][and Fig.\,\ref{Spectra} of this work]{Garufi2020b} and is discussed in Sect.\,\ref{Inner_disk}.

\subsubsection{Azimuthal distribution}
While all continuum maps show azimuthally symmetric structures, the molecular maps show some degree of asymmetry. Similarly to the radial profile of Sect.\,\ref{Radial_distribution}, H$_2$CO and CS show the same behavior. The disks of Haro 6-13 and HL Tau are brighter to the south while that of DG Tau to the north for both molecules. The H$_2$CS detected around HL Tau peaks along the same azimuthal direction as the CS. The CN emission is typically too faint to enable any quantitative analysis but in at least one case, HL Tau, it is azimuthally concentrated at slightly different angles than CS. 

The degree of asymmetry of CS and H$_2$CO also appears comparable. To quantify the asymmetry, we extracted the ratio between the brightest and faintest flux along annuli defined by the geometrical parameters of Table \ref{Disk_properties}. It turns out that in all disks but one, namely, HL Tau, this degree of asymmetry is consistent within the uncertainties for CS and H$_2$CO. Instead, the CS emission around HL Tau is less symmetric than the H$_2$CO emission (a degree of 2.6 versus 1.5).

\subsection{Line brightness and column density} \label{Line_brightness}
The line brightness of {H$_2$CO and CS} was obtained from the velocity-integrated moment 0 maps by integrating over the region determined by the parameters of Table \ref{Disk_properties}. For H$_2$CS (when detected) and CN, {we integrated over a smaller and larger area, respectively,} (see Table \ref{Integrated_fluxes_1}) {to account for the different emitting regions}. The choice of this region does not dramatically impact the total integrated flux but is important to quantify the specific line brightness when constraining the {molecule} column density. The CS and H$_2$CO show a very similar brightness spanning across targets from a ratio of 1.5 to 0.95. When calculated over the same region as CS, the CN shows fluxes from nearly the same to one tenth of the CS. All measurements of the line flux are shown in Appendix \ref{Appendix_fluxes}.

The line brightness can be converted into column density in the scenario of optically thin emission and local thermodynamic equilibrium (LTE). While the latter is a reasonable assumption for the lines targeted in this work, the former may not be. These assumptions are discussed in Sect.\,\ref{Optical_depth}. We constrained the column density of all lines adopting the molecular parameters from CDMS shown in Table \ref{Line_table}. Our poor knowledge of the gaseous temperature was addressed by adopting values between 20 K and 100 K and determining {the range of column densities for this range of temperatures}. Since the H$_2$CO and H$_2$CS lines that we probe are ortho lines, their total column density was obtained by assuming an ortho-to-para ratio of 1.8$-$2.8 \citep{Guzman2018} and 3.0 \citep{Legal2019}, respectively. As for CH$_3$OH, we considered that this molecule exists in so-called A-type and E-type depending on the hydrogen spin and that the ratio between these types is one \citep{Carney2019}. All column densities are listed in Appendix \ref{Appendix_fluxes} and are shown in Fig.\,\ref{Column_densities}.

Generally speaking, the column densities of H$_2$CO and CS calculated for optically thin lines are similar among the different targets, except for {the CS in} Haro 6-13, which shows marginally lower values. On the other hand, the {column} densities of CN are more varied {(up to one order of magnitude)}. In particular, the CN column density in DG Tau and DG Tau B is higher than in the other three objects, and especially HL Tau and Haro 6-13. This dichotomy may, in principle, reflect some different observational settings since DG Tau and DG Tau B were observed at higher resolution. However, the largest recoverable angular scale of this dataset is smaller and therefore the larger CN flux observed is at odds with any possible loss of CN flux at larger scale. 

H$_2$CO exhibits systematically higher column densities than both CS and CN, with their averages of  [7.5, 2.0, 3.0] $\times\ 10^{13}$ cm$^{-2}$, respectively. Figure \ref{Column_densities} also shows that the column density of H$_2$CS where the line is detected, in HL Tau and IRAS04302 \citep[see][]{Codella2020}, is significantly higher than the upper limits put on non-detections. On the other hand, the column density of CH$_3$OH in IRAS04302 \citep[see][]{Podio2020} is still consistent with all non-detections. Interestingly, the targets with detected H$_2$CS (HL Tau and IRAS04302) and CH$_3$OH signal (only IRAS04302) are not those with the strongest CN detected (DG Tau and DG Tau B), {suggesting that the gaseous surface density is not the only determining factor in the  occurrence of the detection of a certain species}.

As is clear from Fig.\,\ref{Column_densities}, approximately a half of the CN hyperfine lines that were probed are undetected. With a few exceptions, all of these non-detections are explained by the lower strength of the relative transition (see Table \ref{Line_table}) since the upper limits that we derived are consistent with the values found from the 226.66 GHz and 226.87 GHz lines. From the diagram, it can also be seen that our upper limits on the HDO and CH$_3$OH lines are not particularly stringent, making it still possible to have any column density up to nearly $10^{14}$ cm$^{-2}$. 


\section{Discussion} \label{Discussion}

\subsection{Line optical depth} \label{Optical_depth}
Any consideration on the spatial distribution and relative abundance of the surveyed molecules cannot ignore the constraints imposed by the optical depth and excitation conditions of their lines. First of all, the column densities that we constrained in Sect.\,\ref{Line_brightness} are only valid in LTE and this equilibrium can be assumed when the critical density of a line is lower than the gas density in the emitting region. Our H$_2$CO, CS, and CN lines all have critical densities lower than $2 \times 10^6$ cm$^{-3}$ in the 20$-$100 K temperature range \citep[e.g.,][]{Shirley2015} and these densities surpass the typical gas densities of a disk only in the uppermost layer \citep[$z/r>0.6$, e.g.,][]{Walsh2010}. Thus, LTE is a reasonable assumption.
 
\begin{figure}
  \centering
 \includegraphics[width=8cm]{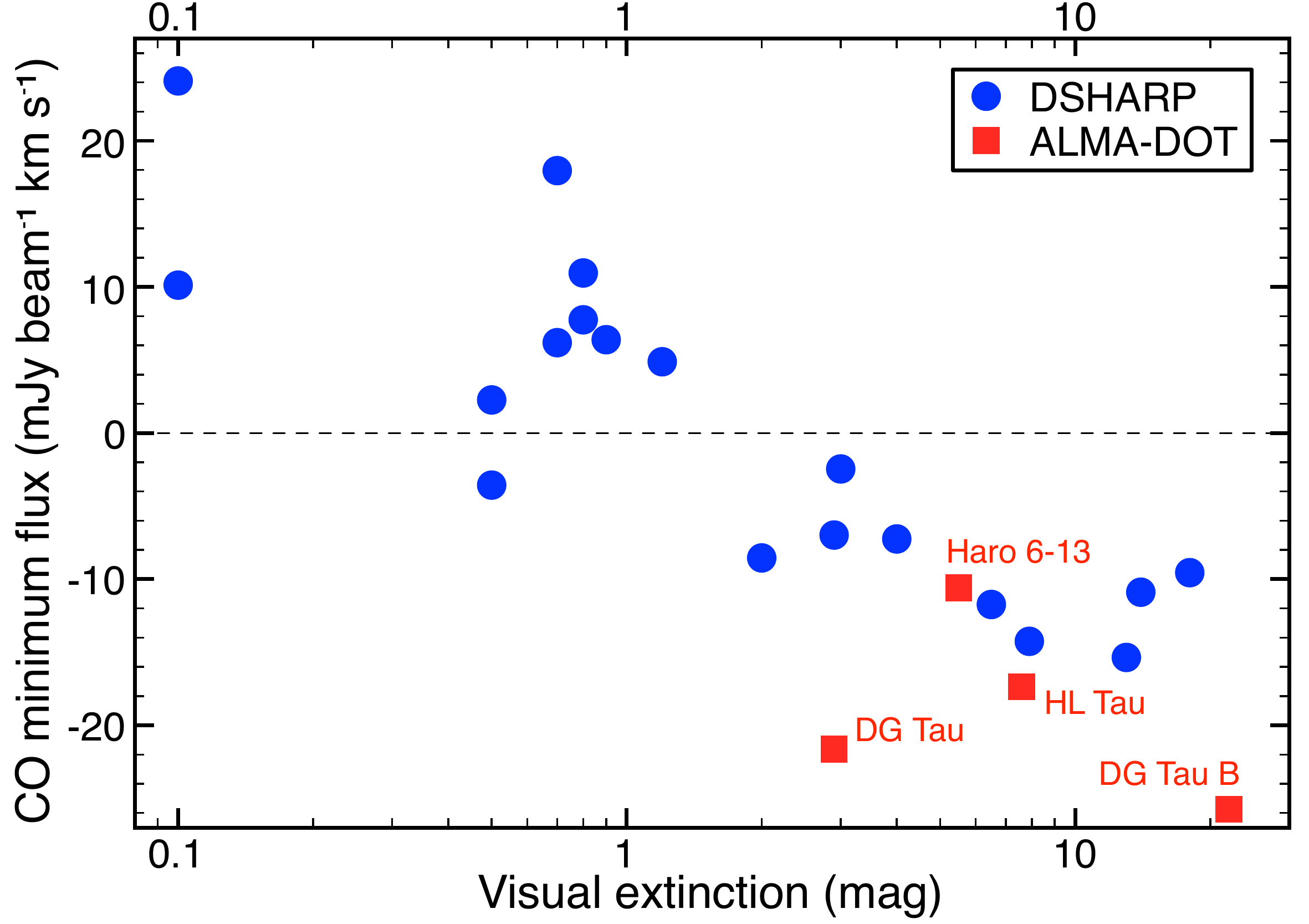} 
     \caption{Minimum flux recorded in the center of the CO continuum-subtracted moment 0 map related to the visual extinction. Fluxes are recalculated after scaling each source to the same distance (140 pc) and beam size (0.1\arcsec$\times$0.1\arcsec). {IRAS04302 is not included because of the edge-on geometry of the disk largely contributing to the high optical extinction.}} 
 \label{Depression_extinction}
 \end{figure}

On the other hand, the line optical thinness is a far more uncertain assumption. In the absence of multiple lines of the same species, we cannot firmly establish whether the lines are optically thin. \citet{Pegues2020} constrained excitation temperature, column density, and line opacity, $\tau_{\rm H_2CO}$, from their H$_2$CO rotational diagram finding that the $\tau_{\rm H_2CO}$ is lower than unity for all their lines and lower than 0.5 for all but one. The H$_2$CO line that we probe was not surveyed by \citet{Pegues2020}. However, their results may suggest that our line is also optically thin {because the upper-level energy (33 K) and strength (43.5 D$^2$) are in the interval of those of their lines (21$-$82 K and 9.1$-$78.3 D$^2$).}

The optical depth of a line can also be tested by comparing its brightness temperature, $T_{\rm B}$, with the gas temperature of the emitting material. In case of optically thick lines, these two quantities should approximately be equal. From the peak of the radial profile of Fig.\,\ref{Radial_profile}, we obtained a peak $T_{\rm B}$ for our targets spanning from 4 to 12 K km s$^{-1}$ and from 5 to 10 K km s$^{-1}$ for CS and H$_2$CO, respectively. The very similar values that we found for CS and H$_2$CO in each object (with discrepancy up to only 2 K) may, in principle, advocate for optically thick lines. However, as noted by \citet{Legal2019}, who obtained comparable values from CS lines, these quantities translate into kinetic temperatures on the order of 7$-$20 K and 8$-$17 K, respectively. Beyond 50 au, these temperatures are typically associated with the disk midplane, whereas the CS and H$_2$CO emission that we probe originates from an upper layer (see Sect.\,\ref{Vertical_profile}). Furthermore, the excitation temperature range that is constrained for CS and H$_2$CO is typically on the order of 25 K \citep{Guzman2018, Legal2019}.  

The discrepancy between brightness and gas temperatures may, in principle, be explained by beam dilution if gaseous substructures or strong gradients on scales significantly smaller than the beam (less than 15 au for DG Tau and DG Tau B) exist. Also, optically thick lines act to absorb the underlying continuum emission at their rest frequency \citep[see e.g.,][]{Weaver2018}. This effect has a direct impact on the measurement of the line brightness. In fact, in the operation of continuum subtraction (see Sect.\,\ref{Observations}), the ground of the line is assumed to be a flat interpolation of the neighboring continuum. The actual level may be lower since the continuum emission is locally absorbed. This may lead to an underestimation of the line brightness and, thus, of $T_{\rm B}$. However, it is unlikely to reconcile the observed $T_{\rm B}$ (<20 K) with the expected gas temperature since the line flux peaks at approximately 100 au where the continuum emission is moderate (see Fig.\,\ref{Radial_profile}). Thus, in absence of better constraints, we assume that the H$_2$CO and CS lines are optically thin.

\subsection{Origin of the inner depression} \label{Inner_disk}

As shown in Sect.\,\ref{Radial_distribution}, the H$_2$CO, CS, and CN fluxes are always dimmed toward the center. This behavior may have several causes, which we discuss here. The first obvious explanation would be a diminished gas surface density, but this is {unlikely given the} full inner extent of the dust.

The peculiarity of this inner region is that it shows negative values around the systemic velocity, $V_{\rm sys}$, and very low values at the line wings (see Fig.\,\ref{Spectra}). Negative values at $V_{\rm sys}$ are best explained by the absorption by foreground material. This effect can act on both the line itself and the local continuum. In turn, the absorption of the local continuum implies an underestimation of the line flux (as explained in Sect.\,\ref{Optical_depth}) that may lead, after the process of continuum subtraction, to negative fluxes where the continuum is stronger (typically in the center). To test whether foreground material is responsible for the dimmed flux at the center, we compared the minimum flux recorded at the center of the CO moment 0 maps with the optical extinction $A_{\rm V}$ retrieved from the literature. We also performed this exercise on the DSHARP sample \citep{Andrews2018} after normalizing each source to the same beam size (0.1\arcsec$\times$0.1\arcsec) and distance (140 pc). A clear trend is shown in Fig.\,\ref{Depression_extinction} and reveals how larger columns of foreground material result in strongly absorbed lines. The ALMA-DOT sources, being very extincted, show severely dimmed fluxes. However, the diagram also suggests that any source with $A_{\rm V}>1$, possibly less, might suffer from this effect.

Nevertheless, the absorption by foreground material cannot explain the paucity of H$_2$CO, CS, and CN flux from the line wings. This flux shortage can only be explained by an actual decrease in emitting molecules or by the aforementioned continuum over-subtraction, that acts to underestimate the line flux. In the disk inner region, the continuum emission away from $V_{\rm sys}$ can be absorbed by optically thick lines, as we discuss in Sect.\,\ref{Optical_depth}. The opposite behavior (continuum shielding lines) is also possible since the molecular layer is closer to the midplane and an optically thick continuum can shield some of the molecular emission. In the latter case, the depression shown by co-spatial molecules should be morphologically the same. 

\begin{figure}
  \centering
 \includegraphics[width=9cm]{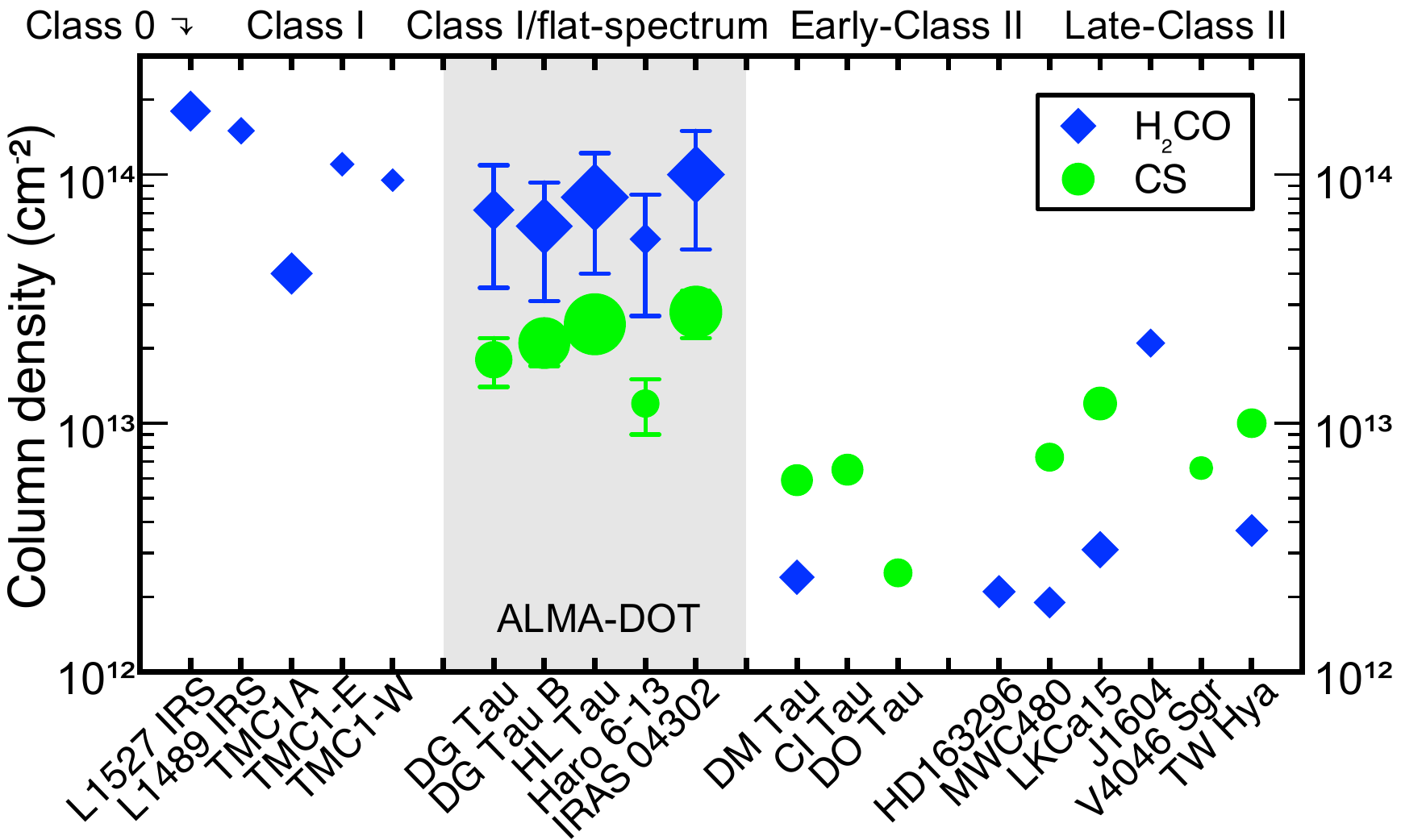} 
     \caption{CS and H$_2$CO column densities from this work compared with the literature. The symbol size reflects the dust disk mass scaling between 5 and 370 M$_\oplus$. {The H$_2$CO estimates of {Class 0 and I} sources are from \citet{vantHoff2020};} HD163296 and TW Hya from \citet{Carney2019}; and all others from \citet{Pegues2020}. All CS estimates are from \citet{Legal2019}, except TW Hya, which is from \citet{Teague2018b}. {Taurus sources (1$-2$ Myr old) are named early-Class II, while the older ones (>2 Myr) are late-Class II.}} 
 \label{Column_literature}
 \end{figure}

All this said, the identical trend shown by the inner CS and H$_2$CO emission around DG Tau B is best explained by the absorption by the continuum, as is also suggested by the fact that the continuum emission is optically thick out to the size of the molecular depression \citep[50 au, see details in][]{Garufi2020b}. On the other hand, CS and H$_2$CO lines behave differently in the other three sources shown in Fig.\,\ref{Radial_profile}. The continuum emission at 1.3 mm around HL Tau is also optically thick in the inner 50 au \citep{Carrasco-Gonzalez2016} but some H$_2$CO flux is detected almost down to the star, similarly to DG Tau. In the latter source, the molecular depression is also displaced from the continuum emission \citep[see][]{Podio2019} and the same effect is visible in HL Tau. This behavior is suggestive of an actual decrease in molecular emission at the inner edge of the gaseous emission (at 30$-$50 au, see Table \ref{Disk_properties}) although the presence of a central flux peak in the stellar proximity (at $10-20$ au) cannot be ruled out by these observations.

\begin{figure*}
  \centering
 \includegraphics[width=17cm]{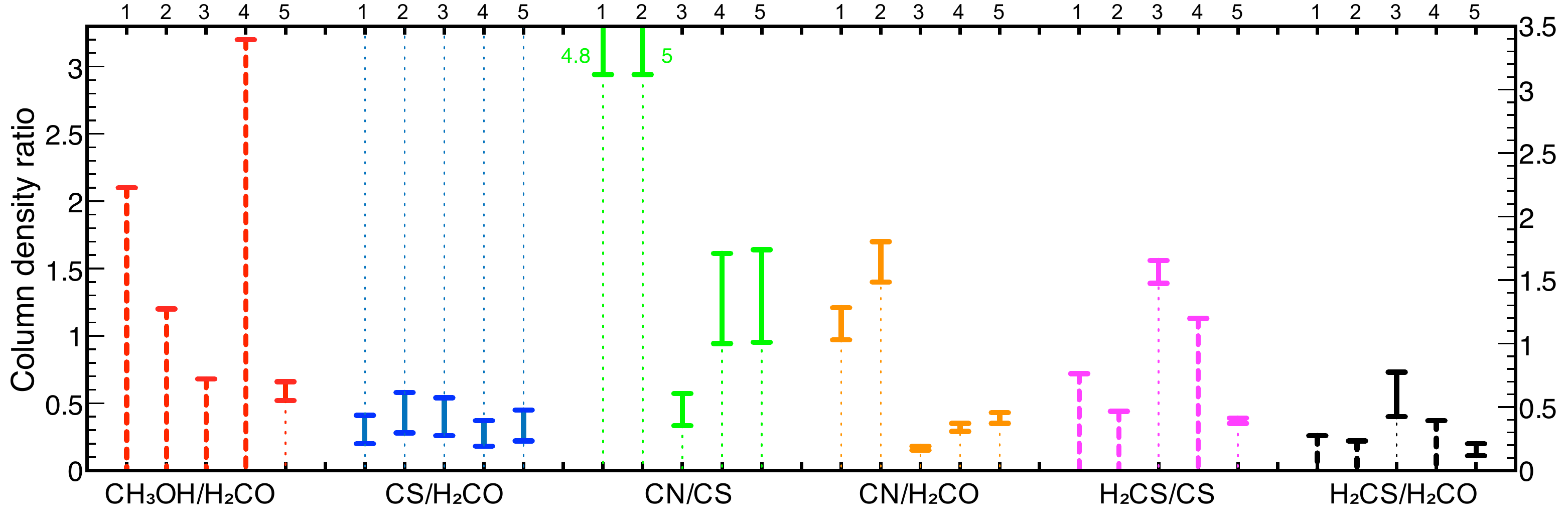} 
     \caption{Column density ratios. Continuous lines indicate the range of solutions found in the hypothesis of optically thin lines in the range of gas temperature 20$-$100 K. Dashed lines indicate upper limits from undetected lines. The possibility that CS and H$_2$CO lines are optically thick is reflected by the thin dotted lines. When a bar falls outside of the diagram, the upper limit is indicated. Sources are ordered as (1) DG Tau, (2) DG Tau B, (3) HL Tau, (4) Haro 6-13, and (5) IRAS04302. } 
 \label{Column_ratio}
 \end{figure*}

\subsection{CN, CS, and H$_2$CO distribution across the disk} \label{Outer_disk}

The observations of the ALMA-DOT survey reveals that CS and H$_2$CO are excellent tracers of the gaseous structure {of Class I disks (with CO contaminated by outflow and envelope emission),} whereas CN is not. This is at odds with \citet{Guilloteau2014}, who used CN as an uncontaminated tracer of the disk dynamical mass. Instead, our maps show that the weak CN emission is only detected at the outer edge of the dust continuum when defined by the 90\% criterion (see Sect.\,\ref{Radial_distribution}). This morphology is consistent with recent thermochemical models predicting CN rings even in full gaseous disks \citep{Cazzoletti2018} and suggesting a dependence between the ring location and the amount of UV emission \citep{vanTerwisga2019}. On the other hand, \citet{Arulanantham2020} concluded that CN is also more readily dissociated in disks with high UV flux, reinforcing the idea that CN substructures are weakly related to the disk physical structure. The column densities we constrained in DG Tau and DG Tau B are approximately consistent with the predictions from \citet{Cazzoletti2018}, while the values found in HL Tau and Haro 6-13 are three to five times lower than what is expected for TTSs.

While a fraction of the CS flux still originates from the extended structures around the disk {\citep[outflow, filaments, see Fig.\,\ref{Overview_kinematics},][2021 in prep.]{Garufi2020b}}, our maps show that the H$_2$CO is best-suited to trace the disk {with no major contamination} even in {these partly} embedded sources. {Nonetheless, \citet{vantHoff2020} showed that some H$_2$CO contamination from the envelope is present for some, possibly younger, sources. Their sample has a target in common with ours, IRAS04302, and the authors also infer the substantial absence of H$_2$CO contamination for this source.} Another possible advantage of H$_2$CO is that multiple transitions in the same ALMA band exist allowing for a rotational diagram analysis. Also, the H$_2$CO likely emits optically thin lines \citep[see Sect.\,\ref{Optical_depth} and][]{Pegues2020} while still being strong enough to map the disk outer regions. \citet{Zhang2020} showed C$^{18}$O, C$^{17}$O, and $^{13}$C$^{18}$O line maps of both HL Tau and DG Tau finding that the first two lines are severely optically thick while the $^{13}$C$^{18}$O is optically thin. However, its integrated flux only amounts to one third of the H$_2$CO line flux measured in this work for the same targets.

Interestingly, \citet{Zhang2020} also determined the gaseous disk extent from $^{13}$C$^{18}$O and found values comparable with the dusty disk size whereas our observations suggest the gas to be 1.6 times more extended when traced by H$_2$CO (see Table \ref{Disk_properties}). Nevertheless, the systematic difference between the gaseous and dusty disk extents constrained in Sect.\,\ref{Radial_distribution} does not necessarily imply dust radial drift following grain growth. Although the difference between the gas and the continuum extent has been used to invoke such an effect \citep[see e.g.,][]{Birnstiel2014, Ansdell2018}, recent thermochemical models show that marginal differences (up to a factor of 4) may entirely or partly be ascribed to the different optical depth of the dust and gas \citep{Facchini2017, Trapman2019}. Thus, the recurrent, small ratio of 1.6 may hint at disks where the dust grain growth and inward drift is moderate and the constancy of the ratio would entirely be due to the different optical depths of dust and H$_2$CO line. The absence of dust drift in these massive Taurus sources would be at odds with the findings by \citet{Trapman2020}, who found that most of the brightest Lupus sources show dust drift. Nonetheless, our sources are younger than the Lupus region (<1 Myr vs 1$-3$ Myr), implying that an earlier evolution of the dust is certainly reasonable.

The overall H$_2$CO and CS morphology revealed by our maps is very similar across all targets. Both molecules appear primarily distributed beyond 50 au in a large ring. Thus far, H$_2$CO has been seen with both centrally peaked and centrally depressed morphologies \citep[e.g.,][]{Oberg2017, Pegues2020}. CS has been mostly observed with centrally peaked morphology by \citet{Legal2019}, although their angular resolution allows the existence of {small ($\lesssim30-40$ au)} depression. Thermochemical disk models \citep[e.g.,][]{Walsh2014, Loomis2015} show that H$_2$CO is efficiently formed both in the gas phase in the warm molecular layer and on the icy mantles of dust grains in the cold outer regions of the midplane outside the CO snowline. Molecules on the icy mantles are then released due to either thermal desorption in the inner and upper disk or non-thermal processes (UV, X-ray, cosmic-ray induced, or reactive desorption). Reasonably, the bulk of the emission observed in our maps lies outside of the CO snowline at the midplane \citep[assumed at 30 au and 50 au in DG Tau and HL Tau,][]{Podio2019, Booth2020}. Therefore, non-thermal desorption of molecules from icy grains is certainly possible, as also suggested by \citet{Podio2020}, who showed that the H$_2$CO flux around IRAS04302 is maximized in the warm molecular layer and is only moderately dimmed (by a factor of 2) toward the midplane.

\subsection{Column densities} \label{Relative_densities}

As we remark in Sect.\,\ref{Line_brightness}, the H$_2$CO column densities are, on average, more than three times higher than the CS column densities. To our knowledge, this work presents the first direct comparison of these two molecules in a sample of disks and putting this result into context is not a straightforward exercise. In Fig.\,\ref{Column_literature}, we compare our H$_2$CO and CS column density with some literature values obtained from moderate and high angular resolution observations. It is clear that our estimates are routinely larger than {those of more evolved sources while they are comparable to those of sources in a similar evolutionary stage \citep[the Class 0/I sample by][]{vantHoff2020}}. {Part} of the observed discrepancy {with more evolved sources could be} due to the larger disk mass of the ALMA-DOT sources (see symbol sizes in Fig.\,\ref{Column_literature}). {However, the young targets by \citet{vantHoff2020} are significantly less massive than the ALMA-DOT sources, possibly suggesting that the discrepancy is related to the different evolutionary stage rather than to the disk mass}. 

Figure \ref{Column_literature} also reveals that the discrepancy between the ALMA-DOT and {the more evolved} sources is more pronounced for the H$_2$CO. In particular, in literature targets with both CS and H$_2$CO column densities available, the former value is always larger. This contrasts with our observations. In principle, this behavior may be explained by the different beam size of the observations if the CS emission is more diffuse than the H$_2$CO emission (as can be seen in our sources, see Fig.\,\ref{Radial_profile}) but it is unlikely to reconcile the observed incongruity. If this is a real effect, it could then be related to the chemical evolution of H$_2$CO and CS in disks, with the former molecule being subject to a more notable evolution. 

The other molecules targeted in this work are also best studied in relation with each other. To compare values that are representative of the same area, we measured all the fluxes used to obtain a useful column density ratio over the same disk region (defined by the parameters in Table \ref{Disk_properties}). The meaningful ratios are shown in Fig.\,\ref{Column_ratio} and listed in Table \ref{Table_ratios}. The range of CS/H$_2$CO ratios across the targets is impressively small (from 0.2 to 0.6), although this could be partly due to the moderately optically thick line emission (see Sect.\,\ref{Optical_depth}). Instead, the CN ratio with both CS and H$_2$CO is very diverse spanning more than one order of magnitude. This diversity must be real as it is inherited from the very different CN line brightness observed (see Sect.\,\ref{Line_brightness}).

The other ratios that we constrained are mostly limits. The column density of H$_2$CS in HL Tau is higher than in the other sources when compared to CS and H$_2$CO. In particular, the H$_2$CS/H$_2$CO ratio can provide information on the S/O ratio if both molecules form {in the gas phase} from the CH$_3$ radical \citep[see e.g.,][]{Fedele2020}. {If both molecules originate from a disk layer where the gas-phase formation dominates, the H$_2$CS/H$_2$CO column density ratio should scale with the S/O ratio.} Our observations show that this ratio is 0.4$-$0.7 in HL Tau, 0.1$-$0.2 in IRAS04302, and <[0.2$-$0.4] in the other sources, {possibly indicating a higher S/O ratio in the disk of HL Tau. However, the line emission ratio also depends on the temperature of the emitting layer as the upper level energy of the two lines is different. Therefore, the derivation of the S/O ratio requires a detailed comparison of the H$_2$CS/H$_2$CO line ratio with the prediction from disk models.} 

Finally, the derived CH$_3$OH/H$_2$CO ratio constrained in IRAS04302 (0.5$-$0.7) is still consistent with all the upper limits of this survey but it is lower than the only ratio ever constrained in a protoplanetary disk \citep[1.3$-$1.7 in TW Hya,][]{Carney2019}. The most meaningful CH$_3$OH/H$_2$CO upper limit that we constrained is that of HL Tau (<0.7), while those of DG Tau (<2.1) and of DG Tau B (<1.2) have been refined from what was published by \citet{Podio2019} and \citet{Garufi2020b}, as shown in Appendix \ref{Appendix_ratios}. {\citet{Podio2020b} also compared the ALMA-DOT CH$_3$OH/H$_2$CO column density ratios with those inferred in the hot corinos around Class 0 sources finding that the latter are between one and two orders of magnitude larger. In the above-cited work, however, we conclude that the current framework does not allow us to disentangle whether these differences are due to a chemical evolution or to different processes involved in their release in the gas phase between hot-corinos (thermal desorption) and disks (non-thermal desorption).}

\subsection{Weak lines} \label{Weak_lines}

Three of the molecular lines that we probed in our campaign are never (HDO), barely (CH$_3$OH), or rarely detected (H$_2$CS). Both H$_2$CS and CH$_3$OH had been imaged for only one protoplanetary disk before \citep[MWC480 and TW Hya, respectively,][]{Legal2019, Walsh2016}, whereas HDO has never been imaged. The H$_2$CS flux that we recovered in HL Tau and IRAS04302 is relatively strong (>5$\sigma$), suggesting that future observations performed with the ALMA-DOT setting (i.e.,\,angular resolution $\lesssim$0.3\arcsec\ and channel-sensitivity $\lesssim$1 mJy, see Appendix \ref{Appendix_fluxes}) will be able to provide more H$_2$CS detections. Beside our $7_{1,6}-6_{1,5}$ line, this may apply to the five existing H$_2$CS lines with strength S$_{ij}\mu^2$>50 D$^2$ and E$\rm _{up}$<100 K, three of which have been probed by \citet{Legal2019}. On the other hand, the CH$_3$OH $5_{0,5}-4_{0,4}$ (A) line remained elusive. We could only recover a 3$\sigma$ detection from the extraordinary disk of IRAS04302 \citep{Podio2020}. While there is no brighter CH$_3$OH lines in the ALMA Band 6, several brighter lines are available in the Band 7. These are three lines with E$\rm _{up}$ comparable to our line (<50 K) and S$_{ij}\mu^2$ from 1.5 to 2.5 times larger, as well as three lines with larger E$\rm _{up}$ (60$-$100 K) and S$_{ij}\mu^2$ from three to four times larger \citep[see also][]{Walsh2016}.

\begin{figure}
  \centering
 \includegraphics[width=9cm]{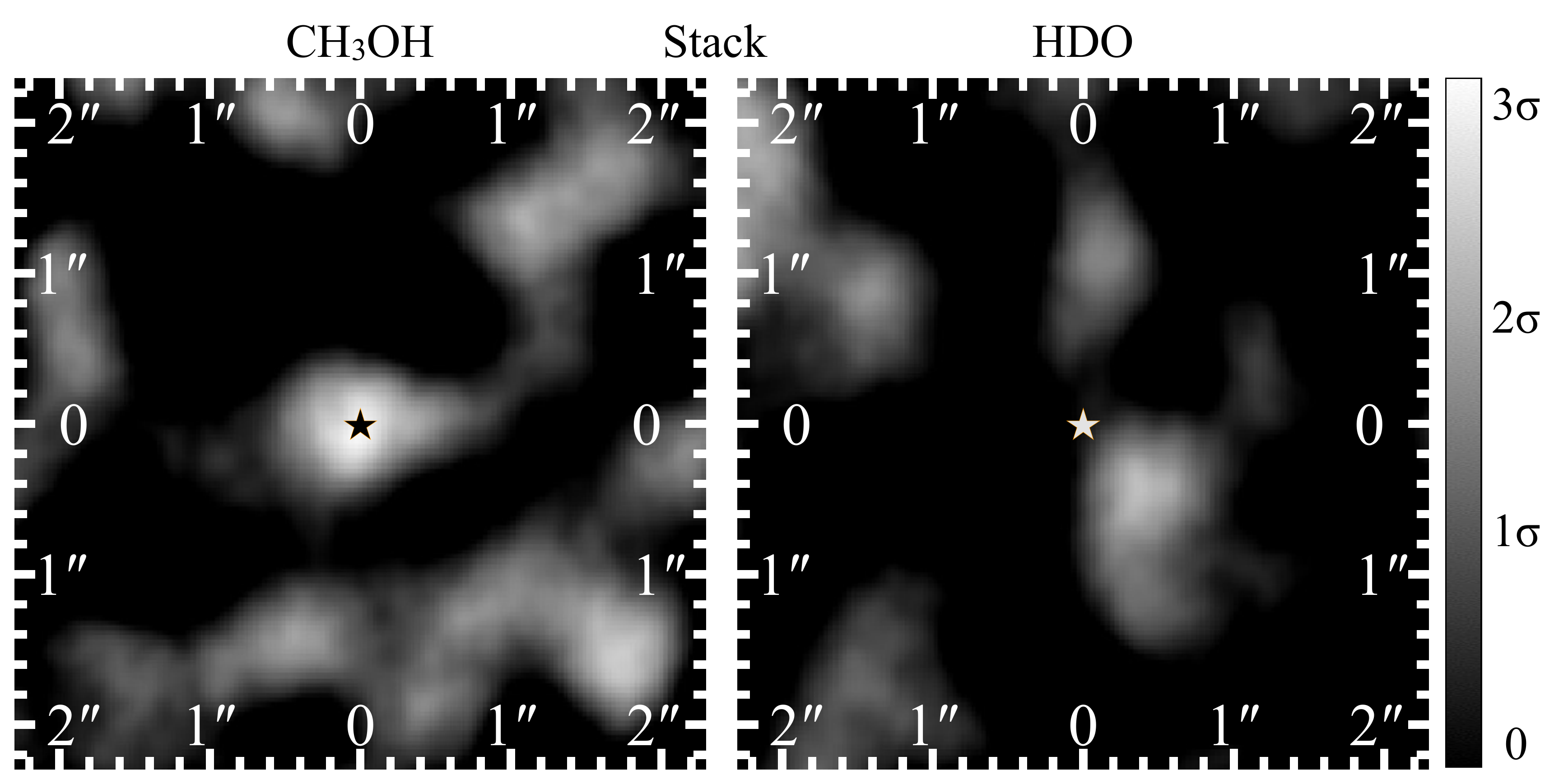} 
     \caption{Moment 0 maps of CH$_3$OH (left) and HDO (right) obtained by stacking maps of all sources with non-detections and smoothing by 1\arcsec.} 
 \label{Undetected_lines}
 \end{figure}

In Fig.\,\ref{Undetected_lines}, we show the stacked map of the undetected CH$_3$OH and HDO lines across all targets, after centering each map on the center of the continuum emission and smoothing the resulting map by 1\arcsec. The CH$_3$OH flux is maximized in the center with a peak at 2.5$\sigma$ suggesting that weak signal is present in (some of) the maps. From this putative stacked detection, we constrained the average column density of CH$_3$OH from DG Tau, DG Tau B, HL Tau, and Haro 6-13 as $[1.5-5]\times10^{13}$ cm$^{-2}$. This value is a half of what is constrained for IRAS04302, suggesting that reasonably deeper observations may yield more detections. The average CH$_3$OH/H$_2$CO column density ratio from this stacked detection would be 0.2$-$0.3. As for HDO, our observations do not set any particular constraints since the upper limits put on the column density are barely comparable to those of the abundant H$_2$CO and CS molecules, and the stacked map shows no putative signal.

\section{Summary and conclusions} \label{Conclusions}
In the context of the ALMA-DOT survey, we probed the emission of ten molecular species from six embedded sources (\mbox{Class I or flat-spectrum}) in Taurus. In this work, we perform a demographical analysis of the spatial distribution in the disks of CS, CN, and H$_2$CO. The main results are: 
\begin{itemize}
\item $^{12}$CO is contaminated by outflow and envelope emission.
\item Both CS and H$_2$CO are good tracers of the gaseous disk having little (the CS) or no (the H$_2$CO) contamination from the surrounding medium.
\item The spatial distribution of the CS and H$_2$CO emission is nearly identical and is also similar across all targets. Both molecules show systematically a strong ring-like emission from large radii (>50 au). The dimmed flux in the inner $\sim$50 au reflects a probable decrease in column density although the presence of an inner flux peak masked by an optically thick continuum in the inner 10$-$20 au cannot be ruled out. 
\item The gaseous disk extent constrained from H$_2$CO is systematically larger by 1.6 than the dust disk. {The constancy and the amount of the ratio suggest} that this is due to the different optical depths of the two components and no radial drifty is indicated.\item CN is not a good proxy of the disk. We routinely observed CN as a faint ring at scales larger than the dust disk. This morphology supports a view where this emission does not trace the disk physical structure but mostly reflects the local UV field.
\end{itemize}

We also constrained the disk-averaged column densities of all available molecules, including H$_2$CS \citep[detected in two disks,][]{Codella2020}, CH$_3$OH \citep[one disk,][]{Podio2020}, and HDO (none). Thanks to the good angular resolution of these observations (from 20 to 40 au), this assessment was performed on smaller areas than in previous works. The main results of this analysis are:

\begin{itemize}
\item The column density of H$_2$CO is routinely larger than that of CS (more than three times larger, on average). This is inconsistent with previous observations of more evolved sources. Their ratio is rather constant across the whole sample (0.2$-$0.6).  
\item Any ratio with CN significantly varies across the targets, as does the CN flux. {Some sources (DG Tau, DG Tau B) show values nearly one order of magnitude higher than the others} (HL Tau, Haro 6-13, and IRAS 04302+2247).
\item The H$_2$CS detected in HL Tau is more abundant than the upper limits set for the other sources when compared to CS (1.5 vs <1). The H$_2$CS/H$_2$CO ratio, which is a possible probe of the S/O ratio, is routinely smaller than 0.7 and is also smaller than 0.4 in all but one source.
\item The average column density of CH$_3$OH from a putative stacked detection would be a half of what is constrained for IRAS 04302+2247. Thus, reasonably deeper observations may yield more detections of CH$_3$OH.
\item The CH$_3$OH/H$_2$CO ratio in IRAS 04302+2247 (0.5$-$0.7) is lower than what was constrained for TW Hya (1.3). Upper limits on the other sources (<0.7 in HL Tau and <1.2 in DG Tau B) also suggest lower values. 
\end{itemize} 

These results set a new benchmark toward the observations of molecular lines from young disks. It is clear that CS and H$_2$CO are uncontaminated and bright enough to trace the gaseous disk structure of embedded and massive disks like those probed by ALMA-DOT. Less exceptional disks will instead require deeper observations. Weaker disk lines such as H$_2$CS and CH$_3$OH remain elusive but our results suggest that slightly deeper observations of similarly bright disks might yield more detections. Broadly speaking, this and the accompanying studies collectively pave the way to future multi-line characterizations of the young Class I disks, enabling the study of the chemical conditions under which the early, and possibly the most important, stages of planet formation occur.

\begin{acknowledgements}
{We thank the referee for fruitful comments.} We are grateful to C.\,Spingola for the help with the ALMA data and to S.\,Perez and D.\,Semenov for the useful discussion. This paper makes use of the following ALMA data: ADS/JAO.ALMA\#2016.1.00846.S and ADS/JAO.ALMA\#2018.1.01037.S. ALMA is a partnership of ESO (representing its member states), NSF (USA) and NINS (Japan), together with NRC (Canada), MOST and ASIAA (Taiwan), and KASI (Republic of Korea), in cooperation with the Republic of Chile. The Joint ALMA Observatory is operated by ESO, AUI/NRAO and NAOJ. This work was supported by the PRIN-INAF 2016 "The Cradle of Life - GENESIS-SKA (General Conditions in Early Planetary Systems for the rise of life with SKA)", the project PRIN-INAF-MAIN-STREAM 2017 "Protoplanetary disks seen through the eyes of new-generation instruments", the program PRIN-MIUR 2015 STARS in the CAOS - Simulation Tools for Astrochemical Reactivity and Spectroscopy in the Cyberinfrastructure for Astrochemical Organic Species (2015F59J3R, MIUR Ministero dell'Istruzione, dell'Universit\`{a}, della Ricerca e della Scuola Normale Superiore), the European Research Council (ERC) under the European Union's Horizon 2020 research and innovation programme, for the Project "The Dawn of Organic Chemistry" (DOC), Grant No 741002, and the European MARIE SKLODOWSKA-CURIE ACTIONS under the European Union's Horizon 2020 research and innovation programme, for the Project "Astro-Chemistry Origins" (ACO), Grant No 811312. We also acknowledge support from INAF/FRONTIERA (Fostering high ResolutiON Technology and Innovation for Exoplanets and Research in Astrophysics) through the "Progetti Premiali" funding scheme of the Italian Ministry of Education, University, and Research, as well as NSF grants AST- 1514670 and NASA NNX16AB48G. This work was also supported by the Italian Ministero dell'Istruzione, Universit\`{a} e Ricerca through the grant Progetti Premiali 2012 - iALMA (CUP C52I13000140001), by the Deutsche Forschungs-Gemeinschaft (DFG, German Research Foundation) - Ref no. FOR 2634/1 TE 1024/1-1, by the DFG cluster of excellence ORIGINS (www.origins-cluster.de), and by the European Union's Horizon2020 research and innovation programme under the Marie Sklodowska-Curie grant agreement No 823823 (RISE DUSTBUSTERS project). This work is also supported by the French National Research Agency in the framework of the Investissements d'Avenir program (ANR-15- IDEX-02), through the funding of the "Origin of Life" project of the Univ. Grenoble-Alpes.
\end{acknowledgements}

\bibliographystyle{aa} 
\bibliography{../MasterReference.bib} 

\begin{appendix}

\section{Observing setup and integrated fluxes} \label{Appendix_fluxes}
Beam size and rms\ of all maps as well as integrated flux and estimated column density of all molecular lines are shown in Tables \ref{Integrated_fluxes_1} and \ref{Integrated_fluxes_2}.

\begin{table*}
 \centering
 \caption{Observing setup, integrated fluxes, and column densities.}
  \label{Integrated_fluxes_1}
  \begin{tabular}{lcccccccc}
  \hline
  Target & Molecule & $\nu_{\rm rest}$ & rms & Beam size & $r_{\rm in}$ & $r_{\rm out}$ & Flux & N$_{\rm X}$ \\
   & & (GHz) & (mJy beam$^{-1}$) & (\arcsec $\times$\arcsec) & (au) & (au) & (mJy km s$^{-1}$) & (10$^{13}$ cm$^{-2}$) \\
  \hline
  \multirow{13}{*}{DG Tau} & H$_2$CO & 225.69777 & $1.3$  & 0.17$\times$0.14 & 35 & 120 & 292 & 3.5$-$11 \\
   & CS & 244.93555 & $0.6$  & 0.16$\times$0.13 & 35 & 120 & 308 & 1.4$-$2.1 \\
   & \multirow{6}{*}{CN} & 226.63219 & $0.9$  & 0.17$\times$0.14 & \multirow{6}{*}{60} & \multirow{6}{*}{180} & < 23 & < 1.7$-$4.3 \\
   & & 226.65956 & $0.9$  & 0.17$\times$0.14 &  & & (36) & (1.4$-$3.6) \\
   & & 226.67931 & $0.9$  & 0.17$\times$0.14 &  & & < 27 & < 1.5$-$3.7 \\
   & & 226.87478 & $1.0$  & 0.14$\times$0.12 &  & & 173 & 2.5$-$6.3 \\
   & & 226.88742 & $1.0$  & 0.14$\times$0.12 &  & & < 28 & < 3.1$-$7.8 \\
   & & 226.89212 & $1.0$  & 0.14$\times$0.12 &  & & < 37 & < 4.1$-$10 \\
   & \multirow{2}{*}{HDO} & 225.89672 & $1.3$  & 0.14$\times$0.12 & \multirow{2}{*}{-} & \multirow{2}{*}{-} & < 11 & < 2.5$-$200 \\
   & & 241.56155 & $0.9$  & 0.13$\times$0.11 &  &  & < 26 & < 5.3$-$26 \\
   & \multirow{2}{*}{CH$_3$OH} & 230.02706 & $1.0$  & 0.16$\times$0.13 & \multirow{2}{*}{-} & \multirow{2}{*}{-} & < 24 & < 20$-$140 \\
   & & 241.79143 & $1.0$  & 0.16$\times$0.12 &  &  & < 24 & < 5$-$21 \\
   & H$_2$CS & 244.04850 & $0.6$  & 0.16$\times$0.13 & - & - & < 19 & < 0.9$-$1.5 \\
   & CO & 230.53800 & 1.0 & 0.14$\times$0.12 & - & - & - & - \\
   \hline
   \multirow{13}{*}{DG Tau B} & H$_2$CO & 225.69777 & $1.4$  & 0.17$\times$0.14 & 45 & 240 & 622 & 3.1$-$9.4 \\
   & CS & 244.93555 & $0.6$  & 0.16$\times$0.13 & 45 & 240 & 920 & 1.7$-$2.5 \\
   & \multirow{6}{*}{CN} & 226.63219 & $0.8$  & 0.17$\times$0.14 & \multirow{6}{*}{150} & \multirow{6}{*}{320} & < 21 & < 0.9$-$3.1 \\
   & & 226.65956 & $0.8$  & 0.17$\times$0.14 &  & & 295 & 2.1$-$5.2 \\
   & & 226.67931 & $0.8$  & 0.17$\times$0.14 &  & & < 29 & < 0.8$-$1.8 \\
   & & 226.87478 & $1.3$  & 0.17$\times$0.14 &  & & 754 & 2.6$-$6.7 \\
   & & 226.88742 & $1.3$  & 0.17$\times$0.14 &  & & < 26 & < 1.2$-$2.9 \\
   & & 226.89212 & $1.3$  & 0.17$\times$0.14 &  & & < 26 & < 1.1$-$2.9 \\
   & \multirow{2}{*}{HDO} & 225.89672 & $0.8$  & 0.17$\times$0.14 & \multirow{2}{*}{-} & \multirow{2}{*}{-} & < 19 & < 2.2$-$185 \\
   & & 241.56155 & $0.9$  & 0.16$\times$0.13 &  &  & < 22 & < 1.8$-$8.1 \\
   & \multirow{2}{*}{CH$_3$OH} & 230.02706 & $0.9$  & 0.17$\times$0.14 & \multirow{2}{*}{-} & \multirow{2}{*}{-} & < 20 & < 17$-$56 \\
   & & 241.79143 & $0.9$  & 0.16$\times$0.13 &  &  & < 26 & < 2.8$-$10 \\
   & H$_2$CS & 244.04850 & $0.6$  & 0.16$\times$0.13 & - & - & < 42 & < 0.6$-$1.1 \\
   & CO & 230.53800 & 1.1 & 0.14$\times$0.12 & - & - & - & - \\
   \hline
   \multirow{13}{*}{HL Tau} & H$_2$CO &225.69777 & $2.0$  & 0.31$\times$0.26 & 55 & 250 & 1130 & 4.0$-$12 \\
   & CS & 244.93555 & $0.9$  & 0.28$\times$0.27 & 55 & 250 & 1553 & 2.0$-$3.0 \\
   & \multirow{6}{*}{CN} & 226.63219 & $2.1$  & 0.32$\times$0.26 & \multirow{6}{*}{180} & \multirow{6}{*}{320} & (81) & (2.4$-$6.0) \\
   & & 226.65956 & $2.1$  & 0.32$\times$0.26 &  & & (70) & (0.6$-$1.5) \\
   & & 226.67931 & $2.1$  & 0.32$\times$0.26 &  & & (77) & (1.7$-$4.4) \\
   & & 226.87478 & $1.8$  & 0.31$\times$0.26 &  & & 180 & 1.0$-$2.4 \\
   & & 226.88742 & $1.8$  & 0.31$\times$0.26 &  & & < 23 & < 1.0$-$2.6 \\
   & & 226.89212 & $1.8$  & 0.31$\times$0.26 &  & & < 22 & < 1.0$-$2.6 \\
   & \multirow{2}{*}{HDO} & 225.89672 & $1.9$  & 0.31$\times$0.26 & \multirow{2}{*}{-} & \multirow{2}{*}{-} & < 39 & < 3.5$-$293 \\
   & & 241.56155 & $2.2$  & 0.30$\times$0.25 & & & < 26 & < 1.6$-$7.1 \\
   & \multirow{2}{*}{CH$_3$OH} & 230.02706 & $2.2$  & 0.31$\times$0.26 & \multirow{2}{*}{-} & \multirow{2}{*}{-} & < 28 & < 24$-$79 \\
   & & 241.79143 & $2.4$  & 0.30$\times$0.25 &  &  & < 26 & < 1.9$-$7.5 \\
   & H$_2$CS & 244.04850 & $0.9$  & 0.28$\times$0.27 & 35 & 190 & 148 & 6.6$-$11 \\
   & CO & 230.53800 & 1.9 & 0.31$\times$0.26 & - & - & - & - \\
    \hline
    \multirow{13}{*}{Haro 6-13} & H$_2$CO &225.69777 & $2.2$  & 0.34$\times$0.26 & 30 & 150 & 206 & 2.7$-$8.3 \\
   & CS & 244.93555 & $0.9$  & 0.32$\times$0.26 & 30 & 150 & 195 & 0.9$-$1.4 \\
   & \multirow{6}{*}{CN} & 226.63219 & $2.0$  & 0.34$\times$0.26 & \multirow{6}{*}{90} & \multirow{6}{*}{180} & < 21 & < 0.7$-$1.8 \\
   & & 226.65956 & $2.0$  & 0.34$\times$0.26 &  & & (51) & (0.5$-$1.3) \\
   & & 226.67931 & $2.0$  & 0.34$\times$0.26 &  & & < 20 & < 0.5$-$1.4 \\
   & & 226.87478 & $2.2$  & 0.34$\times$0.26 &  & & 94 & 0.6$-$1.4 \\
   & & 226.88742 & $2.2$  & 0.34$\times$0.26 &  & & < 23 & < 1.0$-$2.7 \\
   & & 226.89212 & $2.2$  & 0.34$\times$0.26 &  & & < 21 & < 1.1$-$2.7 \\
   & \multirow{2}{*}{HDO} & 225.89672 & $2.1$  & 0.34$\times$0.26 & \multirow{2}{*}{-} & \multirow{2}{*}{-} & < 12 & < 3.6$-$299 \\
   & & 241.56155 & $2.2$  & 0.33$\times$0.24 &  & & < 11 & < 2.5$-$11 \\
   & \multirow{2}{*}{CH$_3$OH} & 230.02706 & $2.0$  & 0.34$\times$0.25 & \multirow{2}{*}{-} & \multirow{2}{*}{-} & < 12 & < 25$-$82 \\
   & & 241.79143 & $1.9$  & 0.32$\times$0.24 &  &  & < 20 & < 5.7$-$23 \\
   & H$_2$CS & 244.04850 & $0.9$  & 0.32$\times$0.26 & - & - & < 19 & < 1.0$-$1.6 \\
   & CO & 230.53800 & 2.0 & 0.34$\times$0.25 & - & - & - & - \\
   \hline
   \end{tabular}
   \tablefoot{Columns: target name, molecular species, line frequency at rest frame, channel rms, synthesized beam size, inner and outer radius used to integrate the flux, integrated flux, and column density calculated in the hypothesis of optically thin lines. The range of column densities reflects the solutions found in the range of gas temperature between 20 K and 100 K. Brackets denote fluxes above 2$\sigma$ uncertainty but below 3$\sigma$. {Upper limits on HDO, CH$_3$OH, and H$_2$CS are measured on the same area as the H$_2$CO and CS.}}
\end{table*}
   
   \begin{table*}
 \centering
 \caption{Observing setup, integrated fluxes, and column densities (continued from \ref{Integrated_fluxes_1}).}
  \label{Integrated_fluxes_2}
  \begin{tabular}{lcccccccc}
  \hline
  Target & Molecule & $\nu_{\rm rest}$ & rms & Beam size & $r_{\rm in}$ & $r_{\rm out}$ & Flux & N$_{\rm X}$ \\
   & & (GHz) & (mJy beam$^{-1}$) & (\arcsec $\times$\arcsec) & (au) & (au) & (mJy km s$^{-1}$) & (10$^{13}$ cm$^{-2}$) \\
  \hline
   \multirow{13}{*}{IRAS 04302+2247} & H$_2$CO &225.69777 & $2.3$  & 0.34$\times$0.26 & 50 & 390 & 2518 & 5.1$-$16 \\
   & CS & 244.93555 & $1.0$  & 0.31$\times$0.26 & 50 & 390 & 2908 & 2.2$-$3.3 \\
   & \multirow{6}{*}{CN} & 226.63219 & $2.2$  & 0.34$\times$0.26 & \multirow{6}{*}{-} & \multirow{6}{*}{480} & 344 & 3.1$-$7.8 \\
   & & 226.65956 & $2.2$  & 0.34$\times$0.26 &  & & 479 & 1.2$-$3.1 \\
   & & 226.67931 & $2.2$  & 0.34$\times$0.26 &  & & 134 & 0.9$-$2.3 \\
   & & 226.87478 & $2.1$  & 0.34$\times$0.26 &  & & 1074 & 1.6$-$4.3 \\
   & & 226.88742 & $2.1$  & 0.34$\times$0.26 &  & & < 41 & < 0.6$-$1.4 \\
   & & 226.89212 & $2.1$  & 0.34$\times$0.26 &  & & < 36 & < 0.6$-$1.4 \\
   & \multirow{2}{*}{HDO} & 225.89672 & $2.3$  & 0.34$\times$0.26 & \multirow{2}{*}{-} & \multirow{2}{*}{-} & < 36 & < 2.0$-$162 \\
   & & 241.56155 & $2.0$  & 0.32$\times$0.24 &  &  & < 29 & < 1.0$-$4.6 \\
   & \multirow{2}{*}{CH$_3$OH} & 230.02706 & $2.1$  & 0.33$\times$0.25 & \multirow{2}{*}{-} & \multirow{2}{*}{-} & < 29 & < 11$-$34 \\
   & & 241.79143 & $2.2$  & 0.32$\times$0.24 &  &  & 64 & 3.8$-$11 \\
   & H$_2$CS & 244.04850 & $1.0$  & 0.31$\times$0.26 & - & 150 & 78 & 7.1$-$12 \\
   & CO & 230.53800 & 2.0 & 0.33$\times$0.25 & - & - & - & - \\
   \hline
   \end{tabular}
\end{table*}

\section{Column density ratio} \label{Appendix_ratios}
The column density ratios obtained in Sect.\,\ref{Relative_densities} and shown in Fig.\,\ref{Column_ratio} are listed in Table \ref{Table_ratios}. The CH$_3$OH/H$_2$CO ratios of DG Tau and DG Tau B are different from what was reported by \citet{Podio2019} and \citet{Garufi2020b} from the same dataset because of the improved data reduction and the formalism adopted in this work to define the emitting region. 

 \begin{table} [H]
 \centering
 \caption{Column density ratios.}
  \label{Table_ratios}
  \begin{tabular}{lccccc}
  \hline
  Ratio & DG & DG B & HL & Haro & IRAS \\
  \hline
  CH$_3$OH/H$_2$CO & <2.1 & <1.2 & <0.7 & <3.2 & 0.5$-$0.7 \\
  CS/H$_2$CO & 0.2-0.4 & 0.3-0.6 & 0.3-0.6 & 0.2-0.4 & 0.2-0.5 \\
  CN/CS & 2.9-4.8 & 2.9-5.0 & 0.3-0.6 & 1.0-1.6 & 1.0-1.6 \\
  CN/H$_2$CO & 1.0-1.2 & 1.4-1.7 & 0.1-0.2 & 0.3-0.4 & 0.3-0.4 \\
  H$_2$CS/CS & <0.7 & <0.4 & 1.4-1.6 & <1.1 & 0.3-0.4 \\
  H$_2$CS/H$_2$CO & <0.3 & <0.2 & 0.4-0.7 & <0.4 & 0.1-0.2 \\
  \hline
  \end{tabular}
\end{table}

\section{Spectral profiles} \label{Spectral_profile}
\begin{figure}
  \centering
 \includegraphics[width=7cm]{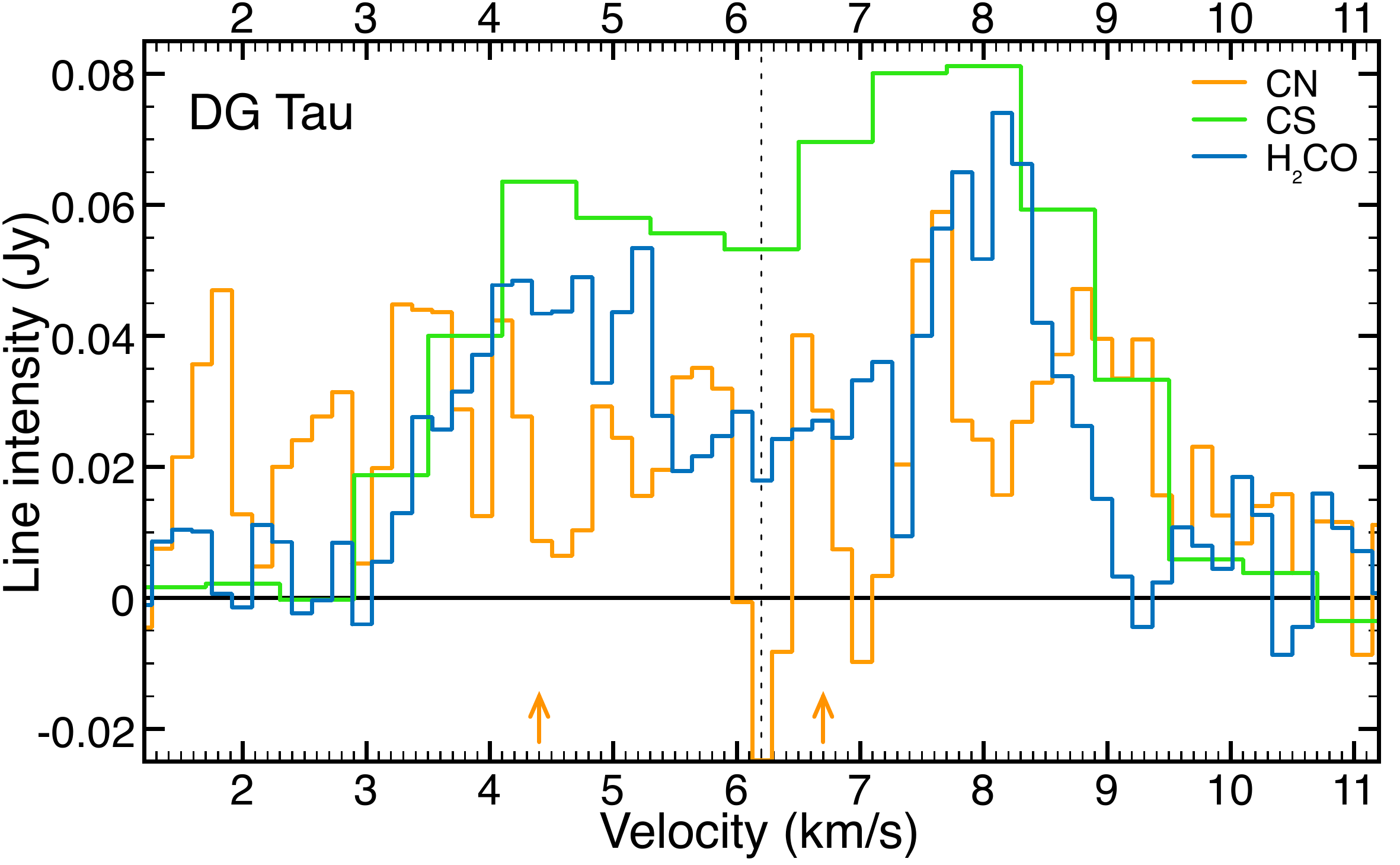}
  \includegraphics[width=7cm]{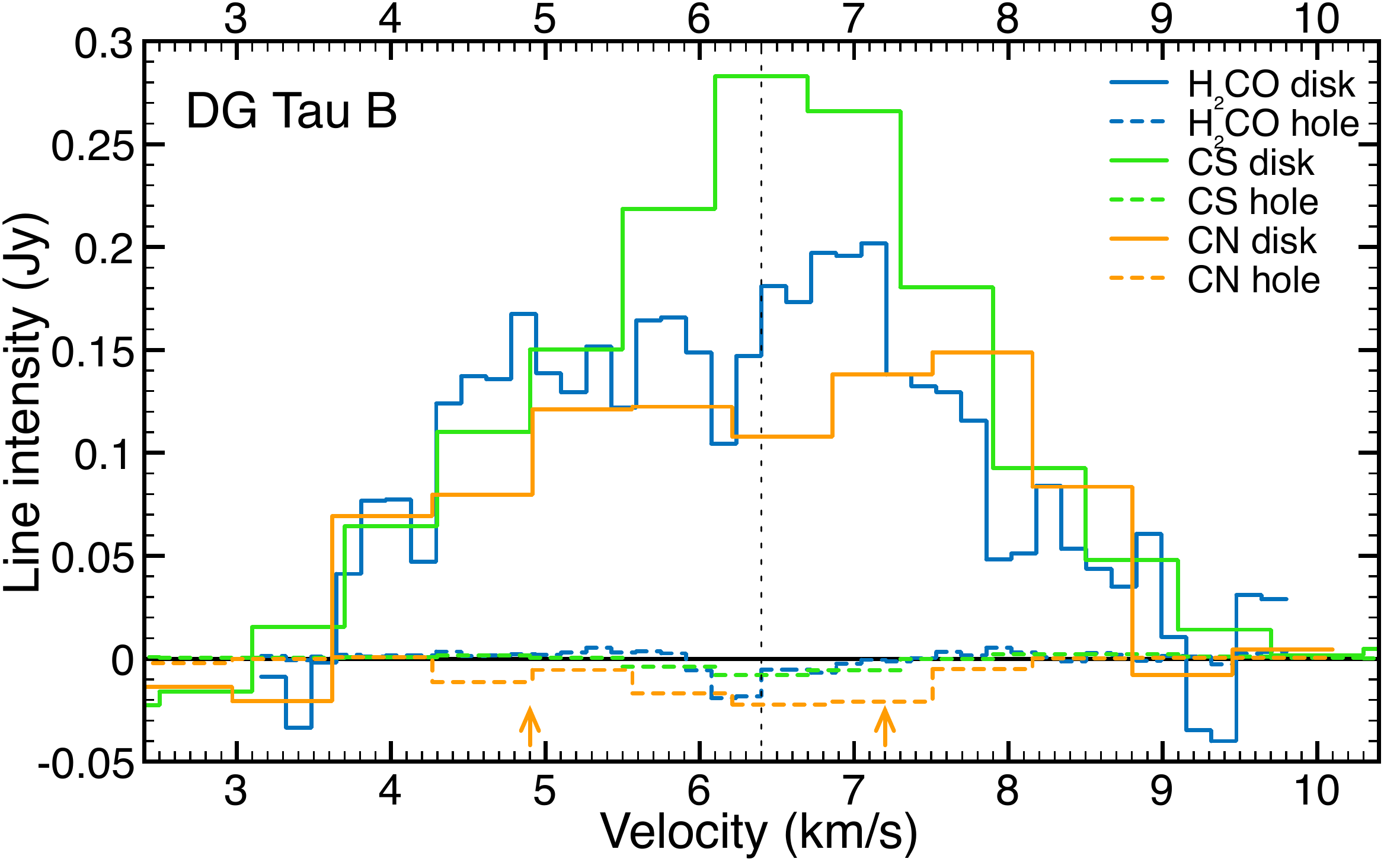}
  \includegraphics[width=7cm]{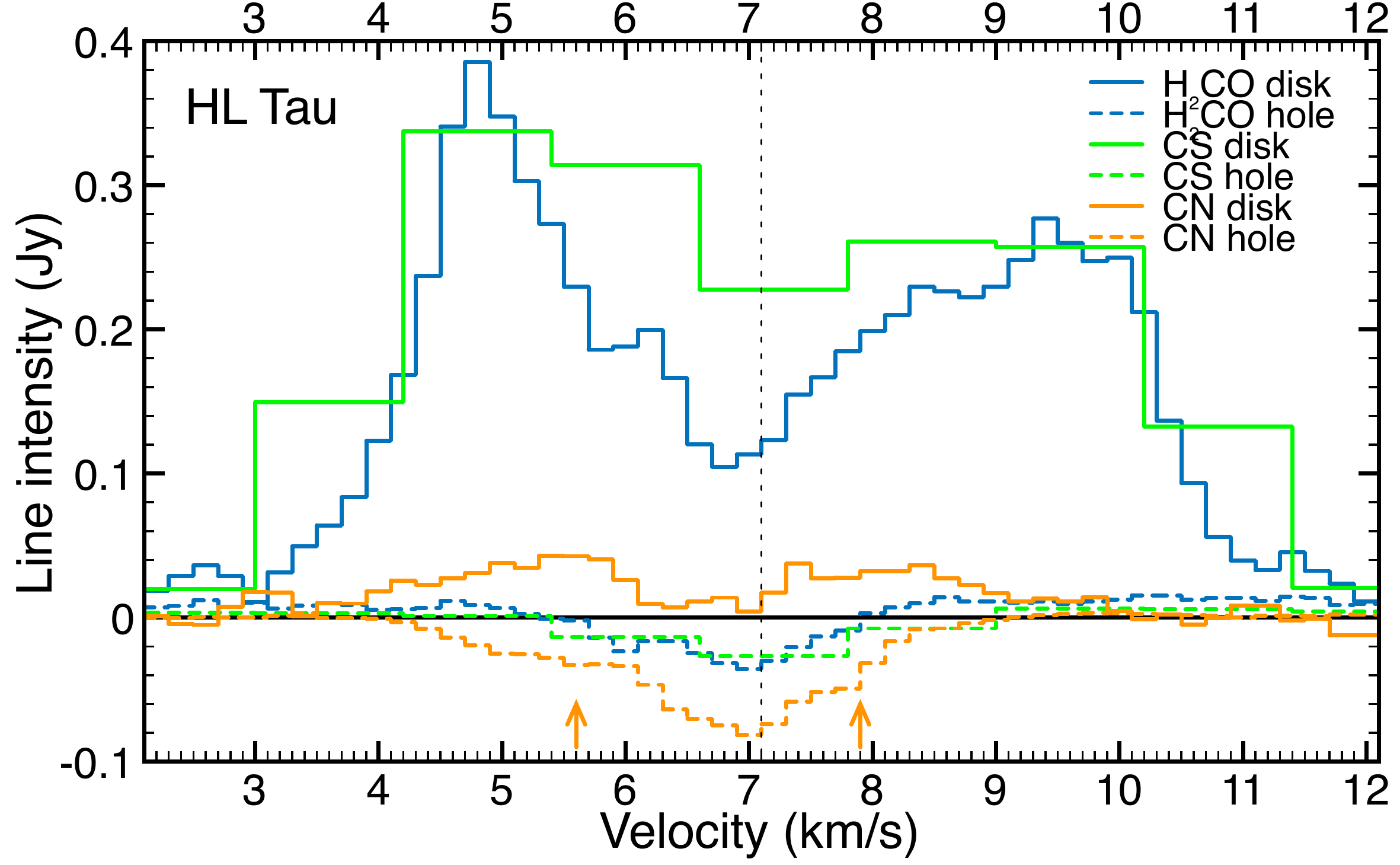}
  \includegraphics[width=7cm]{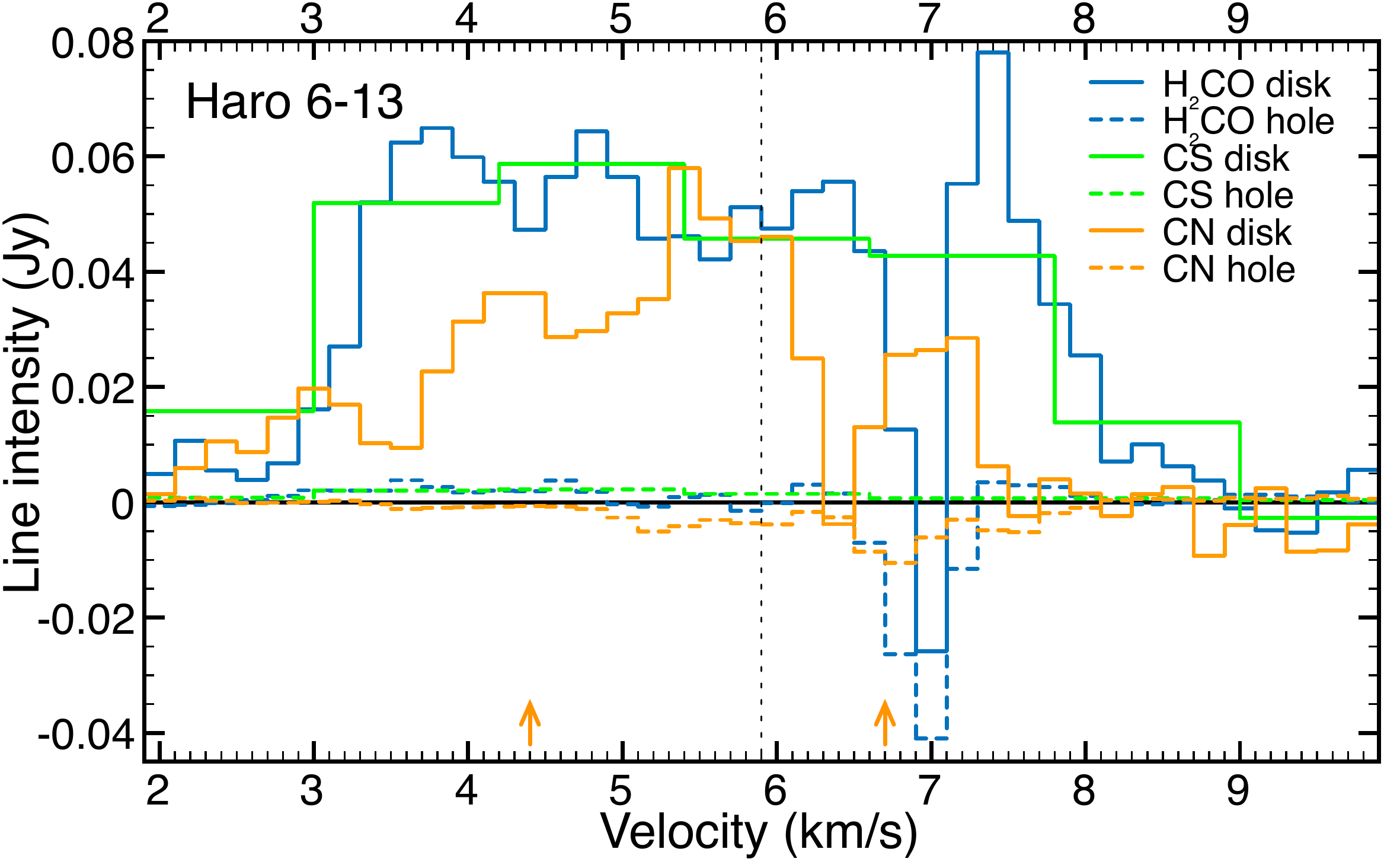}
  \includegraphics[width=7cm]{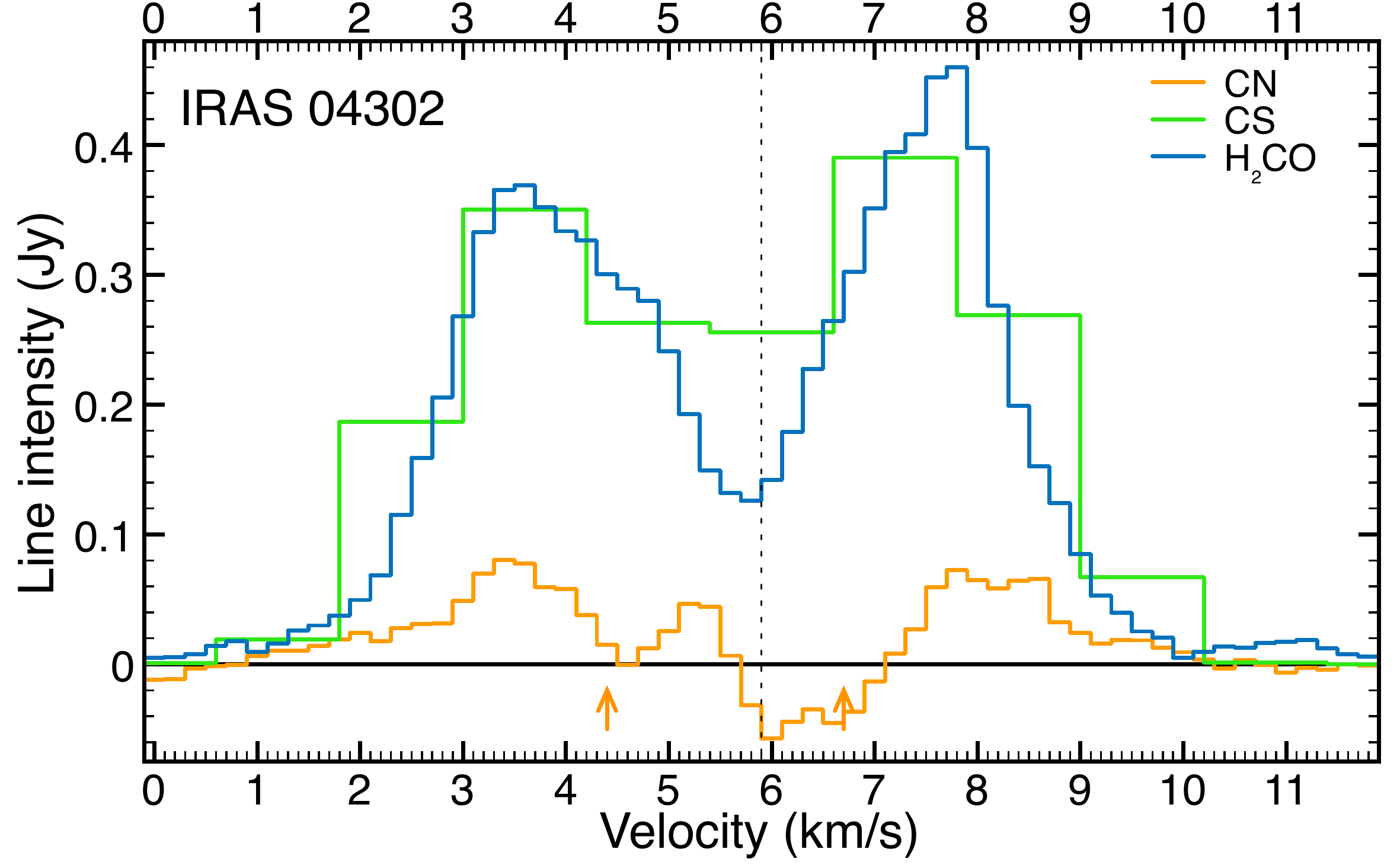} 
     \caption{Spectral profile of H$_2$CO, CS, and CN for {the full sample analyzed in this work}. The profile is obtained over the disk region determined by the parameters of Table \ref{Disk_properties} (continuous lines) and in the inner region with dimmed flux (dashed lines). The vertical line indicates the systemic velocity constrained from the moment 0 map. The two arrows denote the wavelength of the CN blended lines (see Table \ref{Line_table}). Both the horizontal and vertical scales of the {various} diagrams are different.} 
 \label{Spectra}
 \end{figure}

Figure \ref{Spectra} shows the spectral profile of H$_2$CO, CS, and CN for {the targets analyzed in this work}. The main profiles of Fig.\,\ref{Spectra} have been obtained over the disk region determined by the parameters of Table \ref{Disk_properties}. We also extracted the spectral profile of the inner region with depressed flux (see Sect.\,\ref{Inner_disk}). The $V_{\rm sys}$ have been {constrained through the determination of the channel map in which the flux along the disk minor axis and beyond the inner depression (see Sect.\,\ref{Inner_disk}) is maximized. These are 6.2, 6.4, 7.1, 5.9, and 5.9 km s$^{-1}$ for DG Tau, DG Tau B, HL Tau, Haro 6-13, and IRAS04302, respectively}. 

All line spectra of HL Tau show the double-peaked profile typical of rotating, inclined disks. Despite the different spectral resolution of CS and H$_2$CO, their similarity is also clear from this plot. The only difference is that the CS is brighter at the $V_{\rm sys}$ probably because of some material in the immediate surrounding of the disk that is not emitting in the H$_2$CO. The CN line, beside being much fainter, shows peaks closer to the $V_{\rm sys}$. This morphology reflects both the larger extent of the emission and the presence of blended lines (see Table \ref{Line_table}). On the other hand, the negative profile extracted over the inner region is peaked at the $V_{\rm sys}$ for any line (although is slightly more extended over the blueshifted side). This indicates that the material responsible for the absorption of disk emission is at the $V_{\rm sys}$ (see Sect.\,\ref{Inner_disk}).

The spectral profiles of Haro 6-13 are more complex. Overall, the H$_2$CO and CS lines appear flat while the CN line peaks at the $V_{\rm sys}$. Some positive CS flux is detectable even from the inner region, whereas the H$_2$CO is null and the CN flux is negative. Interestingly, the H$_2$CO shows a strong negative component at $\sim +1$ km s$^{-1}$ from $V_{\rm sys}$ {and a bright positive component at $\sim +1.5$ km s$^{-1}$ from $V_{\rm sys}$.} In principle, this feature is reminiscent of a P Cygni profile with the {emission feature redshifted from} the absorption feature. This feature is also visible from the disk emitting area of H$_2$CO, as well as from CN, although a blended line lies close to this velocity.
{More details on DG Tau, DG Tau B, and IRAS04302 can be found in \citet{Podio2019}, \citet{Podio2020b}, \citet{Garufi2020b}, and \citet{Podio2020}.}

\end{appendix}

\end{document}